\documentclass[aps,twocolumn,amsmath,amssymb,groupedaddress,longbibliography]{revtex4-2}
\usepackage{bbm}
\usepackage{mathrsfs}
\usepackage{amsmath}
\usepackage{amsfonts}
\usepackage[colorlinks=true,citecolor=blue,anchorcolor=blue]{hyperref}
\usepackage{graphicx,epstopdf}
\usepackage{subfigure}
\usepackage{epsfig}
\usepackage{dcolumn}
\usepackage{bm}
\usepackage{color}
\usepackage{natbib}
\usepackage{amssymb}
\usepackage{xcolor}
\usepackage{braket}
\usepackage{soul}

\begin{document}

\title{Topological inverse band theory in waveguide quantum electrodynamics}

\author{Yongguan Ke$^{1,2,3}$}

\author{Jiaxuan Huang$^{3}$}

\author{Wenjie Liu$^{3,4}$}

\author{Yuri Kivshar$^{5,6}$} \email{Corresponding author. Email: yuri.kivshar@anu.edu.au}

\author{Chaohong Lee$^{1,2,4,5}$} \email{Corresponding author. Email: chleecn@szu.edu.cn; chleecn@gmail.com}

\affiliation{$^{1}$Institute of Quantum Precision Measurement, State Key Laboratory of Radio Frequency Heterogeneous Integration, Shenzhen University, Shenzhen 518060, China}

\affiliation{$^{2}$College of Physics and Optoelectronic Engineering, Shenzhen University, Shenzhen 518060, China}

\affiliation{$^{3}$Laboratory of Quantum Engineering and Quantum Metrology, School of Physics and Astronomy, Sun Yat-Sen University (Zhuhai Campus), Zhuhai 519082, China}

\affiliation{$^{4}$Quantum Science Center of Guangdong-Hongkong-Macao Greater Bay Area (Guangdong), Shenzhen 518045, China}

\affiliation{$^{5}$Nonlinear Physics Center, Research School of Physics, Australian National University,  Canberra ACT 2601, Australia}

\affiliation{$^{6}$Qingdao Innovation and Development Center, Harbin Engineering University, Qingdao 266000, China}

\date{\today}

\begin{abstract}
	Topological phases play a crucial role in the fundamental physics of light-matter interaction and emerging applications of quantum technologies. However, the topological band theory of waveguide QED systems is known to break down, because the energy bands become disconnected. Here, we introduce a concept of the inverse energy band and explore analytically topological scattering in a waveguide with an array of quantum emitters. We uncover a rich structure of topological phase transitions, symmetric scale-free localization, completely flat bands, and the corresponding dark Wannier states. 
	Although bulk-edge correspondence is partially broken because of radiative decay, we  prove analytically that the scale-free localized states are distributed in a single inverse energy band in the topological phase and in two inverse bands in the trivial phase.  
	Surprisingly, the winding number of the scattering textures depends on both the topological phase of inverse subradiant band and the odevity of the cell number. Our work uncovers the field of the topological inverse bands, and it brings a novel vision to topological phases in light-matter interactions.
	
\end{abstract}

\maketitle

\emph{Introduction.} Light-matter interaction plays a crucial role in the fundamental sciences~\cite{Forn2019,gutzler2021light}, and it underpins the rapid progress of quantum technologies.
Understanding intrinsic mechanisms of absorption as well as spontaneous and stimulated emissions leads to the development of practical applications such as solar cells, LEDs, and lasers~\cite{Atomphoton1998}.
Introducing topology into light-matter interaction could bring new advances such as topological lasers~\cite{st2017lasing,bandres2018topological,harari2018topological,kim2020multipolar}, in which monochromaticity, efficiency, and emission stability become superior to those observed for conventional lasers. 
%
%
To unleash the power of topology in light-matter interaction~\cite{kruk2019nonlinear}, it is important to understand the role of topological phases at the microscopic level of quantum electrodynamics (QED).

Waveguide QED studies photons propagating in waveguides, excitation of quantum emitters, and strong interaction between them~\cite{Lodahl2015,Roy2017, Chang2018,sheremet2021}, providing an excellent platform to explore the interplay between topology and light-matter interaction~\cite{barik2018topological,bello2019unconventional,Downing2019,Luqi2020,Ke2020,Wang2020,Perczel2020,poshakinskiy2021quantum,Leonforte2021,Nie2021,Kim2021,Bernardis2021,Tang2022,jiang2022recent}.
%
%
We notice that the previous studies are based on the well-established topological band theory allowing to calculate topological phases and topological invariants~\cite{Bansil2016,Ozawa2019}.
However, in waveguide QED the energy band splits into two disconnected polariton branches~\cite{Marques2021}, which hinders the straightforward application of the topological band theory.
It is urged to develop a new theoretical approach to explore topological phase in waveguide QED.
Moreover, we need to answer how the bulk topological phase or topological invariant is imprinted in photonic scattering.

\begin{figure}[!h]
	\includegraphics[width=0.48\textwidth]{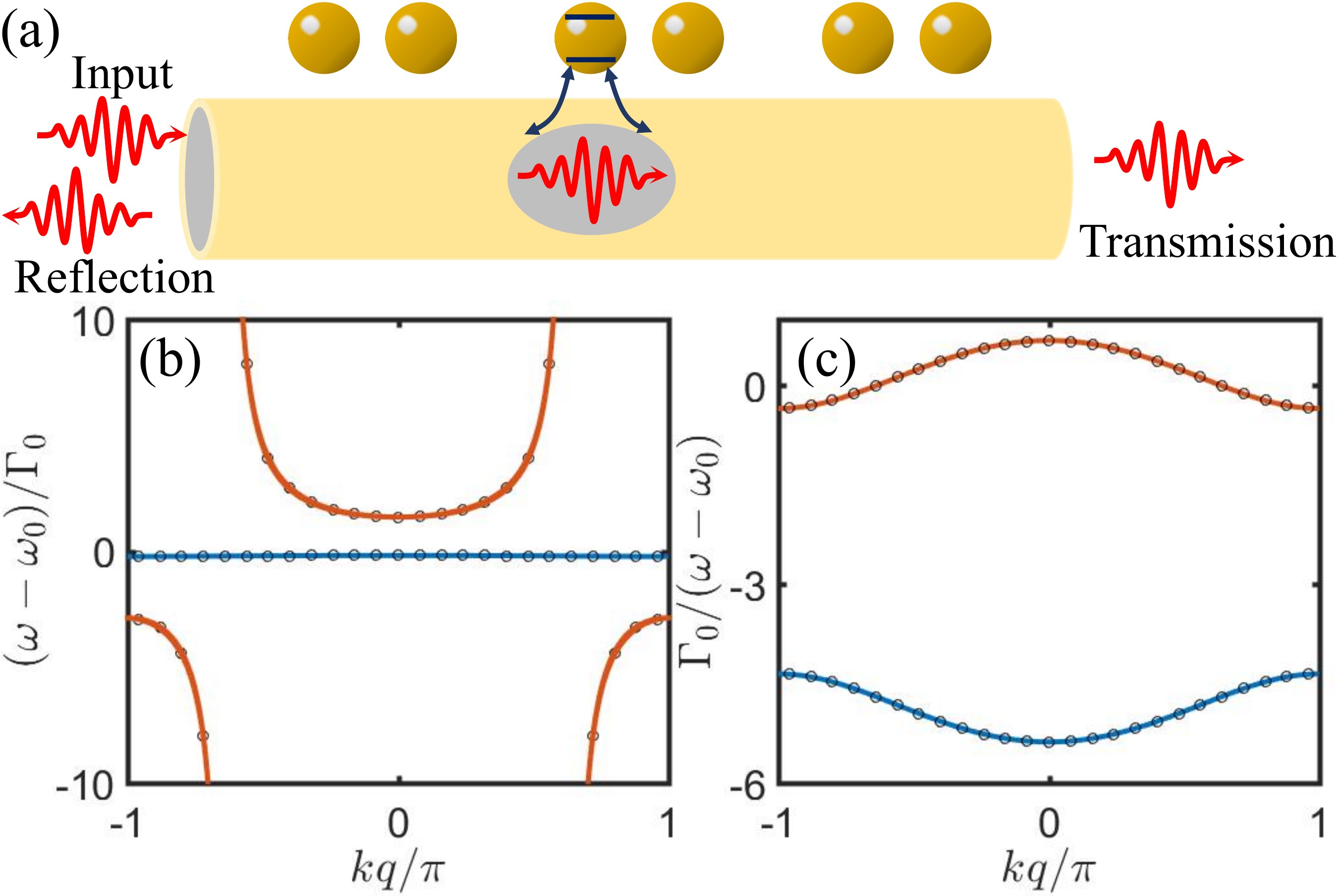}
	\centering
	\caption{Schematic diagram of photonic scattering and inverse energy band.  (a) Diagram of photon in a waveguide scattered by quantum emitters. The emitter positions are arranged as $z_j=d[j+\delta\cos (\pi j +\theta)]$ with the averaged spacing $d$, the modulation strength $\delta$, and the modulation phase $\theta$. (b) Energy band. (c) Inverse energy band. The red and blue inverse energy bands originate from the red and blue bands, respectively. The black circles denote the non-Markov cases in $25$ cells. The parameters are chosen as $\delta=0.4$, $\theta=0$ and $\varphi=1$.
	}\label{Fig1}
\end{figure}

Here, we suggest and implement the topological inverse energy band theory based on a proof-of-principle waveguide QED with inversion symmetry, as depicted in Fig.~\ref{Fig1}, and we uncover a connection between topological phases and photonic scattering.
We find \emph{analytically} unexpected rich topological phase transitions driven by the spatial structure and resonant frequency of emitters.
We also obtain \emph{analytically} the resonant frequencies that favor a completely flat band and a group of dark Wannier states equally occupying two emitters.
%
While the bulk-edge correspondence is partially broken due to radiative decay, we \emph{analytically} prove that scale-free localized states can exist in only one inverse band in topological nontrivial phases and in two inverse bands in trivial phase.
We use the reflection and transmission coefficients to construct scattering textures, which are winding as the photonic frequency is swept along an inverse subradiant band.
Importantly, the winding number of scattering depends on both the topological phase of the inverse band and odevity of the cell number.
We believe our work opens a field of topological inverse band theory, it provides new insights into the role of topological phases imprinted in light-matter interaction, and it paves a way to engineer the dark states for quantum information storage and precision frequency measurements.

\begin{figure*}[!htp]
	\includegraphics[width=0.98\textwidth]{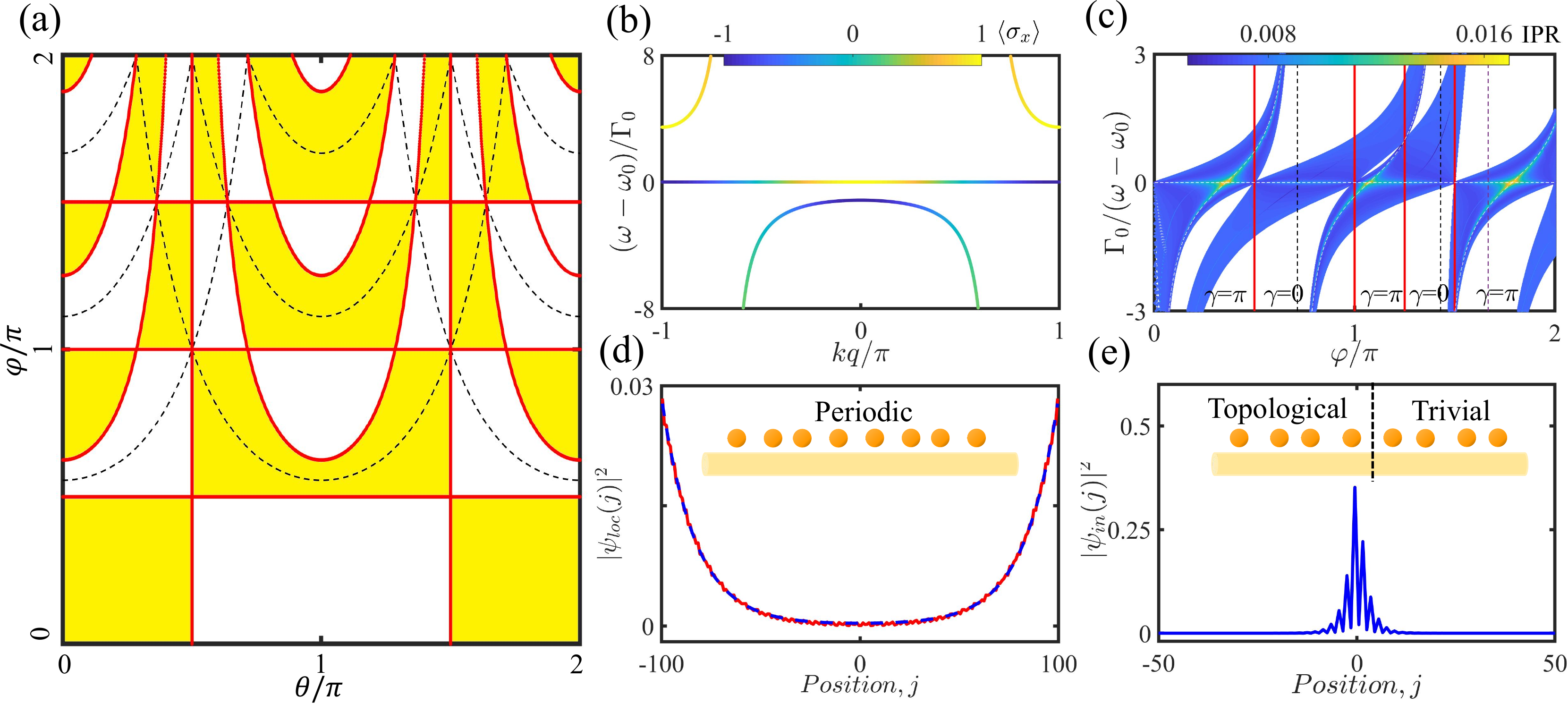}
	\centering
	\caption{Topological phase transition, breakdown of bulk-edge correspondence and topological flat band. (a) Topological phase diagram where the white and yellow regions are trivial and topological phases, respectively. Black dashed lines denote the positions of flat bands. (b) Flat energy band with $\theta=\pi/3$ and $\varphi=5\pi/3$. The colors denote $\langle \sigma_x\rangle$ of the corresponding eigenstates. $\langle \sigma_x\rangle$ are opposite at the high symmetry points, indicating topological phase. (c) The inverse participation ratio of $N=100$ cells under open boundary conditions with $\theta=\pi/3$. There is no edge state at the inverse band gap. white dashed lines denote the centered inverse energy of scale-free localized states. (d) Symmetric scale-free localized states with the largest IPR (red solid line), and the fitting function $|\psi_{loc}(j)|^2 \sim (e^{-\alpha j/N}+e^{\alpha j/N})^2$ (blue dashed line). The other parameters are chosen as $\theta=\pi/3$ and $\varphi=\pi/3$.  (e) Topological interface state in a structure spliced by two arrays with different modulation phases $\theta=\pi/3$ and $\theta=2\pi/3$.     The above calculations are performed with modulation strength $\delta=0.4$.
	}\label{Fig2}
\end{figure*}

\emph{Waveguide QED with inversion symmetry.} As a proof-of-principle example, we consider an array of quantum emitters in a photonic waveguide.
The quantum emitters are two-level systems with resonant frequency $\omega_0$ and positions arranged as $z_j=d[j+\delta \cos (2\pi/q j+\theta)]$; see Fig.~\ref{Fig1}(a).
Here, each unit cell contains $q=2$ emitters, $d$ is the spacing constant, $\delta$ and $\theta$ are the modulation strength and phase, respectively.
The basic process is that the photon is absorbed by emitters and transferred to an excitation, which can be reversely transferred into a photon.
The motion of an excitation is described by an effective Hamiltonian~\cite{Ke2019,Supp}
\begin{eqnarray}\label{effectiveHam}
	H_{eff}=\sum\limits_{j=1}^{2N}\hbar\omega_{0}b_{j}^{\dag}b_{j}^{\vphantom{\dag}}
	-i \hbar \Gamma_0 \sum\limits_{j,l=1}^{2N} b_j^\dag b_l e^{i\varphi|z_l-z_j|/d}.
\end{eqnarray}
Here, $N$ is the number of cells, and the scaled Planck constant $\hbar$ is set as a unit in the following.
$\Gamma_0=g^2/c$ is the radiative decay rate of a single emitter with $g$ denoting the coupling strength between excitation and photon and $c$ being the velocity of light.
The hopping of excitation from any two emitters is assisted by emission and reabsorption of a photon in the waveguide.
The hopping strength is an infinite range with a phase shift, in which the phase unit $\varphi=\omega d/c$ depends on the injected photonic frequency $\omega$.
When the spacing constant is small enough [$d\ll c/(\omega-\omega_0)/N\sim \lambda/(2\pi N)$], one may take the Markov approximation by replacing the phase constant with $\varphi=\omega_0 d/c$~\cite{Ke2019}.

By applying Bloch theorem, we can immediately obtain the Hamiltonian in momentum space $H_k$ and the Bloch states $|{\psi _{n,k}}\rangle  = \sum_{j,l}^{} {{e^{i  k q j}}{u_{n,k}(l)}|q j + l\rangle }$ with energy $\omega_{n,k}$~\cite{Supp}.
We find that the energy band is diverged at $k=\pm \varphi$ and is split into upper and lower polariton branches; see Fig.~\ref{Fig1}(b).
The divergence of the energy band originates from the infinite-range coupling in the Markov approximation.
When considering the non-Markov case in a small system with $N=25$ cells, there is no divergence in the non-Markov energy bands~\cite{Supp}; see the black circles in Fig.~\ref{Fig1}(b).
However, the disconnection of energy bands in both Markov and non-Markov cases hinders the application of topological band theory.
To avoid the disconnection, we define inverse energy band as
\begin{equation}
	\bar \omega_k=(\omega_k-\omega_0)^{-1},
\end{equation}
which can be obtained by solving $H_k ^{-1}|\bar u_{n,k}\rangle=\Gamma_0 \bar \omega_{n,k} |\bar u_{n,k}\rangle$. Here, the amplitudes of state $|\bar u_{n,k}\rangle$ is given by $\bar u_{n,k}(l)$ which are the same as $u_{n,k}(l)$ if $n$ and $\bar n$ are one-to-one correspondence. 
The inverse energy band glues the upper and lower polariton branches and becomes a continuous function; see solid lines for Markov case and black circles for the non-Markov case in Fig.~\ref{Fig1}(c).
By fixing the disconnection problem, we can easily define topological phases and topological invariants based on the inverse band as usual.
We find that the energy bands, the inverse energy bands, and the topological phase in Markov and non-Markov cases are consistent with each other~\cite{Supp}. 
Although the Hamiltonian and its inversion share the same eigenstates, they have different band indices and thus the inverse energy band may uncover novel topological states which are failed to be found by the original energy band.

\emph{Topological phase transition, partial breakdown of bulk-edge correspondence and dark Wannier states.} Since the effective Hamiltonian has the inversion symmetry $H(i,j)=H(2N+1-i,2N+1-j)$, the Zak phase~\cite{Zak1989}, as a geometrical phase picked up by a particle sweeping the Brillouin zone, is quantized. The Zak phase of the $n$th inverse band is given by,
\begin{equation}
	\gamma=i\int_{-\pi/2}^{+\pi/2} \langle \bar u_{n,k}|\frac{\partial}{\partial k} |\bar u_{n,k}\rangle dk.
\end{equation}
Fig.~\ref{Fig2}(a) shows the Zak phase as a function of the modulation phase and the hopping phase constant.
%
%
Surprisingly, the topological phase diagram is unexpectedly rich, with a nontrivial phase $\gamma=\pi$ for yellow regions and a trivial phase $\gamma=0$ for white regions.
The red lines and the black dashed lines mark the topological phase boundaries, and the parameters for the flat band with energy $\omega=\omega_0$~\cite{Supp}.
When a flat band with energy $\omega=\omega_0$ exists, because its inverse band becomes divergent, we have to combine the energy band to extract its topological phase; see Fig.~\ref{Fig2}(b) for the topological case with $\delta=0.4$, $\theta=\pi/3$ and $\varphi=5\pi/3$.
For the continuous flat band, $\langle \sigma_x\rangle=\langle u_{n,k}|\sigma_x|u_{n,k}\rangle$ changes from $-1$ at $k=\pi/2$ to $1$ at $k=0$, where $\sigma_x$ is the Pauli matrix.
The sign difference of $\langle \sigma_x\rangle$ between high-symmetric points is related to the Zak phase, similar to Ref.~\cite{Fu2017}.
For a fixed spatial structure, we can drive the topological phase transition by the resonant frequency $(\varphi=\omega_0/cd)$, which can be controlled by either the resonant frequency of an injected photon or the Zeeman shift induced by magnetic fields.
This is in stark contrast to the topological phases in electronic materials which are largely determined by the crystal structure.

To explore the bulk-edge correspondence, we calculate the inverse participation ratio of eigenstates in a finite array under open boundary conditions, $\text{IPR}=\sum |\psi_n(j)|^4$, where $\psi_n(j)$ is the amplitude of the $n$th eigenstate at the $j$th site.
The $\text{IPR}$ tends to $1$ for the most localized state and $0$ for the extended state, which can be used to distinguish localized and extended states.
Fig.~\ref{Fig2}(c) shows the dependence of IPRs on the hopping phase constant and the inverse energy.
The red solid lines separate the energy spectrum into several regions with different Zak phases.
There are no edge modes in the inverse band gaps, regardless of trivial or topological phases.
The breakdown of the conventional bulk-edge correspondence can be explained by the inverse of the effective Hamiltonian, which turns out to be an Su-Schrieffer-Heeger (SSH) model with effective radiative defects at the boundaries~\cite{Supp}.
It immediately becomes clear that an excitation at the end sites may well escape away from the system of emitters as a propagating photon.

However, the bulk topological phase still has significant consequences.
First, symmetric scale-free localized states [Fig.~\ref{Fig2}(d)], $|\psi_{loc}(j)|^2 \sim (e^{-\alpha j/N}+e^{\alpha j/N})^2$, exist in the continuous spectrum with larger IPRs~\cite{Supp,li2021impurity,li2023scalefree}.
Surprisingly, the distribution of scale-free localized states in inverse energy bands depends on topological properties; see Fig.~\ref{Fig2}(c).
In the topological phase with $\gamma=\pi$, the scale-free localized states can be distributed in only one single inverse energy band; in the trivial phase with $\gamma=0$,  they can be distributed in both two inverse energy bands.
We analytically prove this observation and obtain the centered inverse energy of the scale-free localized states denoted by the white dashed lines in Fig.~\ref{Fig2}(c)~\cite{Supp}.
Second, the number $(S)$ of subradiant states in the energy band around $\omega_0$ is $S=N$ for a trivial phase and $S=N-1$ for a topological phase. 
These subradiant states become dark states when the band becomes completely flat with energy $\omega=\omega_0$~\cite{Supp}.
Because the flat band has zero group velocity and infinite effective mass, the lifetime of excitation can be infinitely long.
In the flat band, we find $N$ dark Wannier states being inter-cell superposition $(|2j-1\rangle+|2j\rangle)/\sqrt{2}$ ($j=1,2,...,N$) in the trivial phase and $N-1$ dark Wannier states being intra-cell superposition $(|2j\rangle+|2j+1\rangle)/\sqrt{2}$ ($j=1,2,...,N-1$) in the topological phase.
These dark Wannier states form a decoherence-free subspace, which is simple, arranged in order, and with the same energy of resonant frequency $\omega_0$.
These features could potentially be used for quantum memory, quantum information process and precision frequency measurement~\cite{Supp}.
Third, we can further extract the Zak phase of the inverse energy band via long-time average of mean cell position in quantum walks~\cite{Longhi18,Jiao2021,Supp}.

%
%
%

%
%

We need to emphasize that the radiative defects at the boundaries only partially break the bulk-edge correspondence.
In a splicing structure which connects two arrays with different Zak phases, we find topological interface states [Fig~\ref{Fig2}(e)] appearing in the inverse energy gaps.  
Because of no radiative defects at the interface, the topological interface states in large inverse energy gap are immune to disorder to some extent.
In contrast to the topological materials, in the hybrid quantum waveguide QED systems the bulk-edge correspondence is preserved in the interface between two arrays but fails in the interface between an array and a vacuum.

\begin{figure}[!htp]
	\includegraphics[width=0.48\textwidth]{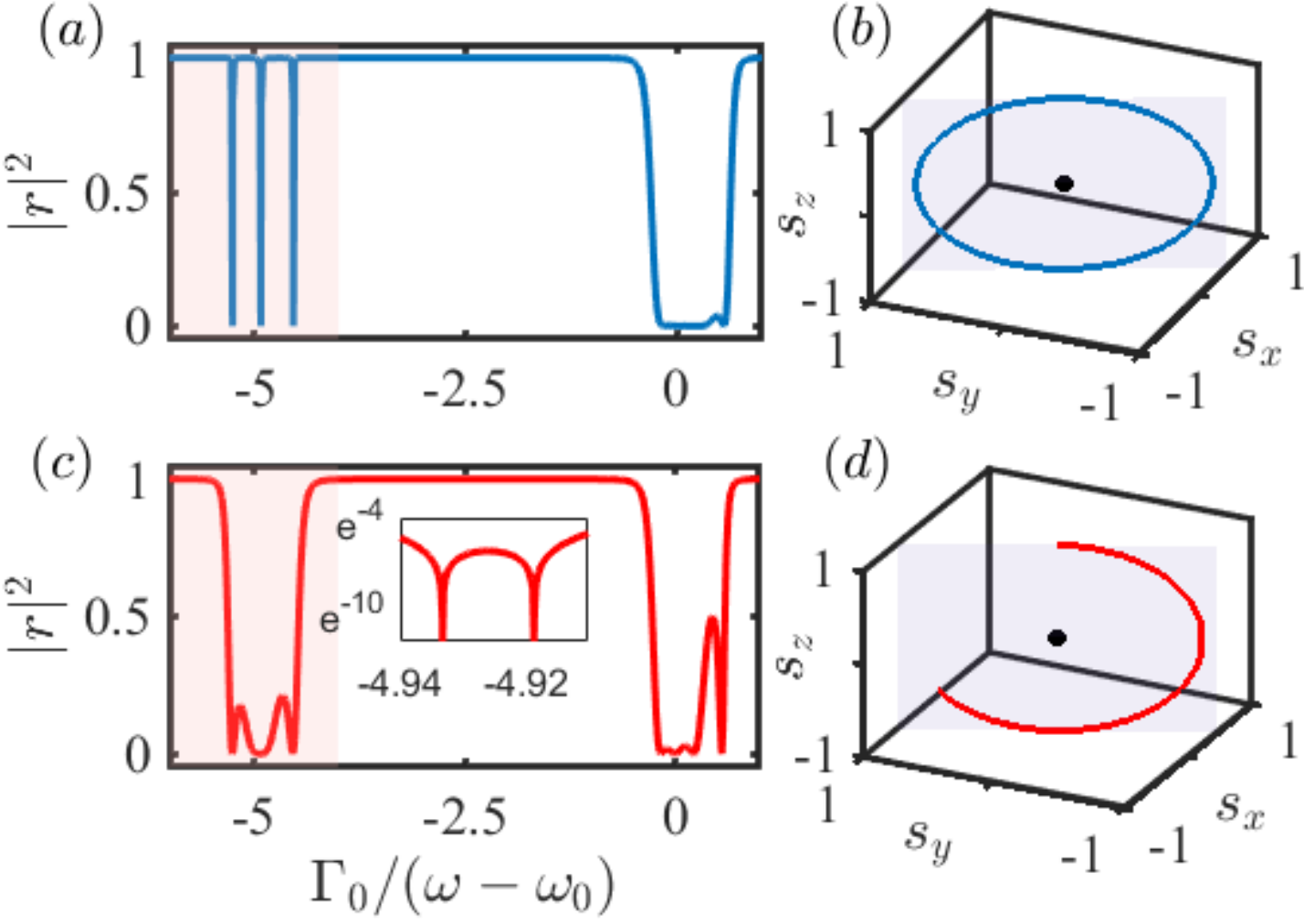}
	\centering
	\caption{Topology-dependent scattering in four cells. (a,c) Reflection as a function of inverse energy in topological and trivial phases, respectively. The inset of (d) enlarges its region with two nearby dips.  There are $N (N-1)$ dips in reflection spectrum around the subradiant inverse energy band for trivial (topological) phase. (b,d) The trajectory of $(s_x, s_y, s_z)$ as the inverse energy sweeps through the shading regions for topological and trivial phases, respectively. The parameters are chosen as $N=4$, $\delta=0.4$, $g=1$, $\Gamma_0=0.01$, $\omega_0=c/d=100$, $\theta=0$ and $\theta=\pi$ for topological and trivial phases, respectively.
	}\label{Fig3}
\end{figure}
%
\emph{Bulk topology of photonic scattering.} Because an excitation will generally decay into a photon in the long-time dynamics, it is highly appealing to explore how the topology of excitation is mapped to the topology of photonic scattering.
We consider a photon with momentum $(\kappa)$ injected into the waveguide. After interacting with the emitters and transferring into excitation, or vice versa, the photon is either reflected or transmitted.
%
%
The reflection and transmission coefficients are given by
\begin{equation}
	r_{{{\kappa}}}=-i\Gamma_0 \sum_{j,j'}G_{j,j'}(\omega_{{{\kappa}}})e^{i\omega_{{{\kappa}}}/c (z_j+z_{j'})}, \label{reflectionk}
\end{equation}
and
\begin{equation}
	t_{{{\kappa}}} = 1 - i{\Gamma _0}\sum\limits_{j,j'}^{} {{{G_{j,j'}}(\omega_{\kappa} ) e^{i\omega_{\kappa} /c({z_{j'}} - {z_j})}}}, \label{transmissionk}
\end{equation}
which satisfy $|r_{{{\kappa}}}|^2+|t_{{{\kappa}}}|^2=1$~\cite{Supp}. Here, $G(\omega)=(\omega-H_{eff})^{-1}$ is the Green's function for an excitation.
%
%
%
%
We calculate the reflection spectrum as a function of the inverse energy for the cases of  topological nontrivial and trivial phases; see Figs.~\ref{Fig3}(a) and \ref{Fig3}(c), respectively.
%
%
When the photonic frequency is in resonance with the subradiant states, the photon is completely transmitted, and there are $N-1$ ($N$) dips for the topological nontrivial (trivial) phase in the shadow regions where the photonic frequency sweeps across the lower inverse band.

To show how the topological phase affects the scattering, we need to calculate the scattering textures $ s_i =\langle \psi_\kappa|\sigma_i|\psi_\kappa\rangle~(i=x,y,z)$, where $\{|\psi_{\kappa}\rangle=(r_\kappa,t_\kappa)^{T}\}$ and $\sigma_i$ are Pauli matrices.
The scattering textures can potentially be probed by interference between reflected and transmitted photon.
With the scattering textures $\vec{\bold{s}} =(s_x,s_y,s_z)$, we can define the winding number of scattering along the $x (y)$ direction as
\begin{equation}
	\nu_{x(y)}=\int \big[\vec{\bold {s}} \times \frac{\partial \vec{\bold{s}}}{\partial \bar \omega}\big]_{x(y)} d\bar \omega.
\end{equation}
where $\bar \omega$ sweeps through the lower inverse band.
%
%
%
However, $\nu_{x(y)}$ is generally not quantized along the $x(y)$ directions.
As the trajectories $(s_x,s_y,s_z)$ in Figs.~\ref{Fig3}(b) and \ref{Fig3}(d) locate in a plane parallel to the $z$ direction and depart from both the $x$ and $y$ directions,
we can define a winding number around original point and along the direction perpendicular to the plane as
\begin{equation}
	\nu=\sqrt{\nu_x^2+\nu_y^2}.
\end{equation}
We summarize the winding number in different topological phases with even and odd cells; see Table~\ref{tab}.
%
\begin{table}[t!]
	\begin{tabular}{c|c|c}
		\hline\hline
		& Topological nontrivial & Topological trivial \\ \hline
		Even cell & 1 & 0  \\ \hline
		Odd cell & 0 & 1  \\ \hline \hline
	\end{tabular}
	\caption{Winding number for the  trajectory of scattering textures~\cite{Supp}.}\label{tab}
\end{table}
%
Remarkably, the winding number of the scattering textures depends on both the topological phase and the odevity of the cell numbers.
We numerically find that a subradiant state contributes a $\pi$ phase shift~\cite{Supp}.
Based upon the fact that there are $(N-1)$ subradiant states in a topological phase and $N$ subradiant states in a trivial phase,
we can reasonably argue that both the topological nontrivial phase with even cells and the trivial phase with odd cells contribute $\pi$ phase to the scattering texture, leading to nontrivial quantization of winding number $\nu$.

\begin{figure}[!htp]
	\includegraphics[width=0.49\textwidth]{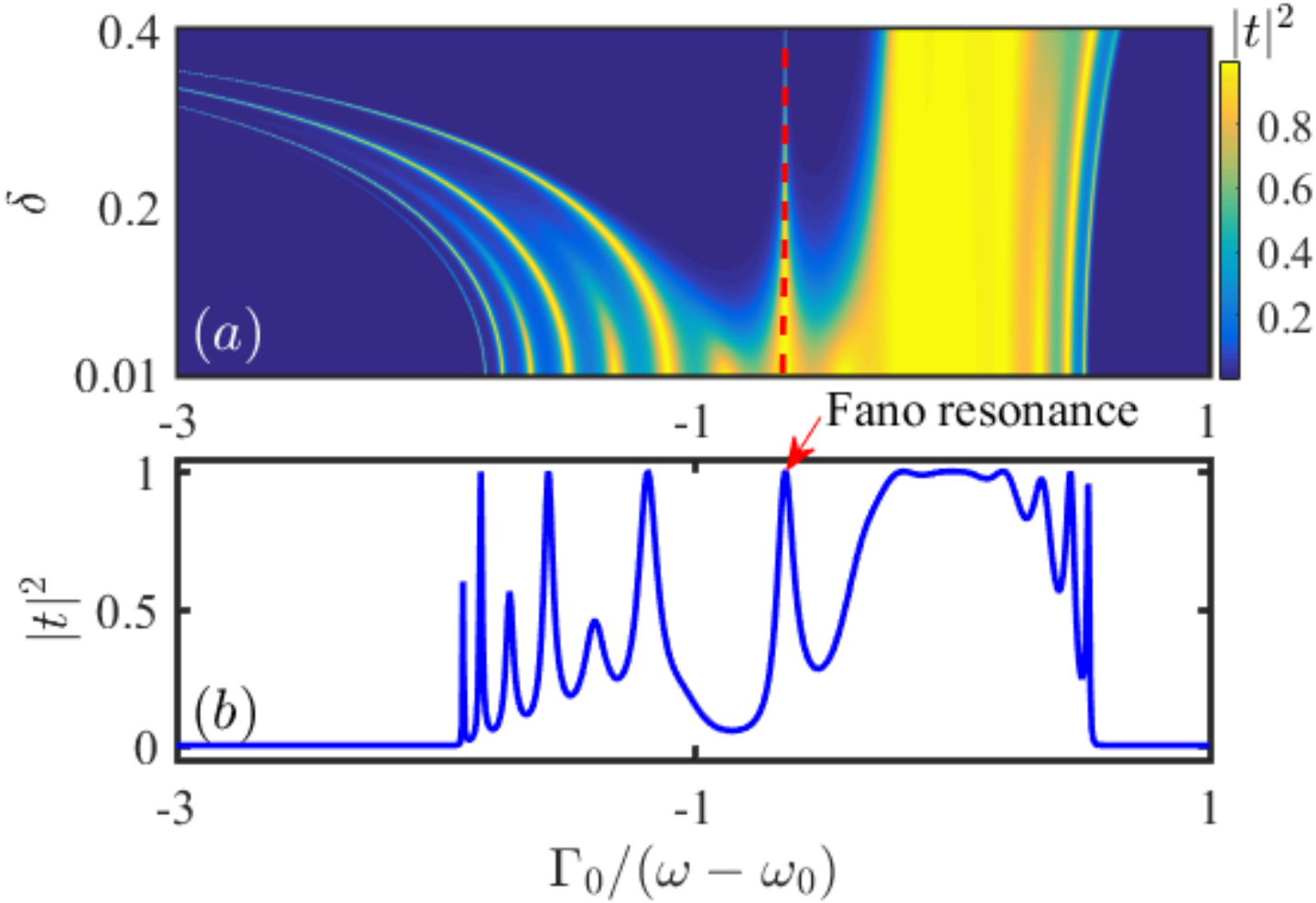}
	\centering
	\caption{(a) Transmission enhanced by topological interface state and (b) Fano resonance due to the topological interface state in the case of $\delta=0.1$. The parameters are chosen as $N=8$, $g=1$, $\Gamma_0=0.01$, $\omega_0=c/d=100$. The modulation phases of the two connecting arrays  are given by $\theta=0$ and $\theta=\pi$.
	}\label{Fig4}
\end{figure}

\emph{Topological interface state in transmission spectrum.} 
Topological interface state exists in the structure spliced by two arrays with different topological properties. Because the topological interface state is a subradiant state, it will enhance the transmission when an injected photon is in resonance with it.
We show the transmission spectrum as a function of the modulation strength $\delta$ and the inverse frequency; see Fig.~\ref{Fig4}(a).
The red dashed line denotes the inverse frequency of the topological interface state in the inverse band gap, at which the transmission is strongly enhanced. 
At a small modulation strength $\delta=0.1$, the profile of the transmission spectrum becomes an asymmetric Fano resonance around the topological interface state~\cite{limonov2017fano,Zangeneh2019}.
The Fano resonance is an interplay between the topological interface state and the states close to the inverse band edge.
%

%
%
%
\emph{Discussion.} Since the inverse energy band is related to the inverse of Hamiltonian known as Green's function, the topological inverse energy band actually reflects the topological phase of Green's function, which is directly related to the scattering of a photon.
It means that the topological inverse band could be a powerful and natural theoretical framework for predicting and tailoring novel topological-enriched light-matter interactions.
We expect an unexplored richness in topological QED within the topological inverse band theory, considering higher dimensions that favor a variety of symmetries~\cite{Bernardis2021,vega2022topological}, more photons that support stimulated emission, quantum correlations and entanglement~\cite{Ke2020} and cavity mediated long-range interactions~~\cite{PhysRevX.8.011002,Chanda2021selforganized,mivehvar2021cavity}.

\begin{acknowledgments}
	We acknowledge useful discussions with Alexander Poddubny, Ling Lin, Li Zhang, and Linhu Li. This work is supported by the National Key Research and Development Program of China (Grant No. 2022YFA1404104), the National Natural Science Foundation of China (Grant No. 12025509 and 12275365), the Key-Area Research and Development Program of Guangdong Province (Grant No. 2019B030330001), and the Natural Science Foundation of Guangdong Province (Grant No. 2023A1515012099).
	Y. Kivshar is supported by the Australian Research  Council (Grants DP200101168 and DP210101292).
\end{acknowledgments}




%

\onecolumngrid
\clearpage

\renewcommand {\Im}{\mathop\mathrm{Im}\nolimits}
\renewcommand {\Re}{\mathop\mathrm{Re}\nolimits}
\renewcommand {\i}{{\rm i}}
\renewcommand {\phi}{{\varphi}}

\begin{center}
	\noindent\textbf{\large{Supplemental Materials:}}
	\\\bigskip
	\noindent\textbf{\large{Topological inverse band theory in waveguide quantum electrodynamics}}
	\\\bigskip
	\onecolumngrid
	
	Yongguan Ke$^{1,2,3}$, Jiaxuan Huang$^{3}$, Wenjie Liu$^{3,4}$,  Yuri Kivshar$^{5,6,*}$, Chaohong Lee$^{1,2,4,5 \dag}$
	

\small{$^{1}$Institute of Quantum Precision Measurement, State Key Laboratory of Radio Frequency Heterogeneous Integration, Shenzhen University, Shenzhen 518060, China} \\
\small{$^{2}$College of Physics and Optoelectronic Engineering, Shenzhen University, Shenzhen 518060, China}\\
\small{$^{3}$Laboratory of Quantum Engineering and Quantum Metrology, School of Physics and Astronomy, Sun Yat-Sen University (Zhuhai Campus), Zhuhai 519082, China}\\
				\small{$^{4}$Quantum Science Center of Guangdong-Hongkong-Macao Greater Bay Area (Guangdong), Shenzhen 518045, China}\\
	\small{$^{5}$Nonlinear Physics Center, Research School of Physics, Australian National University,  Canberra ACT 2601, Australia}\\
	\small{$^{6}$Qingdao Innovation and Development Center, Harbin Engineering University, Qingdao 266000, China}
\end{center}

\setcounter{equation}{0}

\setcounter{figure}{0}

\setcounter{table}{0}
\renewcommand{\theequation}{S\arabic{equation}}

\renewcommand{\thefigure}{S\arabic{figure}}
\renewcommand{\theHfigure}{S\arabic{figure}}

\renewcommand{\thesection}{S\arabic{section}}


\section{Effective Hamiltonian for the excitation}
We consider the propagation of photons in a waveguide coupled to two-level atoms with quasiperiodic spacing, which is characterized by the Hamiltonian in the real space that contains three parts~\cite{Shen2007S},
\begin{equation}
	H_{R} = {H_A} + {H_F} + {H_I}. \label{RealSpaceHam}
\end{equation}
Here, $H_A$ describes the energy of excitations,
\begin{equation}
	{H_A} = \sum\limits_j^{} {\hbar {\omega _0}b_j^\dag } {b_j}
\end{equation}
where $\omega_0$ is the uniform resonant frequency of the two levels in atoms,
${b_j^{\dag}}$ ($b_j$) is the excitation creation (annihilation) operator on the $j$th atom.
${H_F}$ describes the propagating photon in the waveguide along the $z$ direction with light velocity $c$,
\begin{equation}
	{H_F} = i\hbar c\int_{ - \infty }^{ + \infty } {dz \big[ {a_L^\dag (z)\frac{\partial }{{\partial z}}{a_L}(z) - a_R^\dag (z)\frac{\partial }{{\partial z}}{a_R}(z)} \big]},
\end{equation}
where $a_L^\dag(z)$ ($a_L(z)$) and $a_R^\dag(z)$ ($a_R(z)$) are the creation (annihilation) operators for a photon propagating leftward and rightward at position $z$, respectively.
${H_I}$ describes the interaction between excitation and photon,
\begin{equation}
	{H_I} = \sum\limits_{j=1}^{N} {\hbar g\int_{}^{} {dz\delta (z - {z_j})\left[ {\big( {a_L^\dag (z) + a_R^\dag (z)} \big){b_j} + h.c.} \right]} },
\end{equation}
where $g$ is the coupling strength, and the total number of atoms is $N$. 
The coupling between photon and excitation happens at the positions of atoms, which are given by
\begin{equation}
	z_j=d[j+\delta \cos (2\pi\beta j+\theta)],
\end{equation}
where $d$ is the spacing constant, $\delta$ is the periodic strength, $\beta$ is the modulation frequency and $\theta$ is the modulation phase.
Atomic positions become quasiperiodic when $\beta$ is an irrational number. Here, we only consider $\beta=1/2$, indicating that a unit cell contains two atoms.

Alternatively, by making a Fourier transform,
\begin{eqnarray}
	a_L(z)&=&\frac{1}{\sqrt{2\pi}}\int_{-\infty}^0 a(k) e^{i kz}dk,\nonumber  \\
	a_R(z)&=&\frac{1}{\sqrt{2\pi}} \int_{0}^{+\infty} a(k) e^{i kz}dk,
\end{eqnarray}
we can derive the Hamiltonian in the momentum space,
\begin{eqnarray}\label{eq:H0}
	H_{M}&=&\int_{-\infty}^{+\infty} \hbar\omega_{k}a(k)^{\dag}a(k)^{\vphantom{\dag}}dk
	+\sum\limits_{j}\hbar\omega_{0}b_{j}^{\dag}b_{j}^{\vphantom{\dag}} \\
	&+&\frac{\hbar g}{\sqrt{2\pi}}\sum\limits_{j}\int_{-\infty}^{+\infty}
	(b_{j}^{\dag}a{(k)}e^{i k z_{j}}+b_{j}a{(k)}^{\dag}e^{-i k z_{j}})dk.\nonumber
\end{eqnarray}
Here, $a{(k)}$ ($a{(k)}^{\dag}$) is the annihilation (creation) operator for a photon with a wave vector $k$.
$\omega_k=c|k|$ is the frequency of a single photon with light velocity $c$.

Similarly to Ref.~\cite{Ke2019S}, by tracing the degree of photon, the motion of an excitation in the atomic array is governed by
\begin{equation}
	H_{eff} = \sum_j{\omega_0 b_j^{\dagger}b_j}-i\Gamma_0 \sum_{j,j'} { e^{i\omega/c|z_j-z_{j'}|}b_j^{\dagger}b_{j'}}.
\end{equation}
Here,  $\Gamma_0=g^2/c$ is the radiative decay rate, and the phase constant is defined as $\varphi=\omega d/c$, depending on the frequency of an input photon $\omega$, spacing constant $d$ and the light velocity $c$.
The hopping of excitation from $z_l$ to $z_j$ points is assisted by the photon emitted from the $l$th atom and consequently reabsorbed by the $j$th atom.
When the spacing constant is small enough, the phase constant in the effective Hamiltonian can be replaced by $\varphi=\omega_0 d/c$, which is the so-called Markov approximation~\cite{Ke2019S}.

\section{Energy band, inverse energy band and Zak phase}
\subsection{Bloch Hamiltonian in momentum space}
In this subsection, we derive the Bloch Hamiltonian for calculating the energy band in the momentum space.
The position of the $j$th atom is given by
\begin{equation}
	z_j = d\left[j+\delta \cos{(2\pi\beta j+\theta)} \right].
\end{equation}
For arbitrary $m$, $n$, we obtain the difference between $z_m$ and $z_n$,
\begin{equation}
	\frac{z_m-z_n}{d} = m-n-2\delta \sin{(2\pi\beta \frac{m+n}{2}+\theta)}\sin{(2\pi\beta \frac{m-n}{2})}.
	\label{Diff}
\end{equation}
We use the cell index $j$ and the sublattice index $l$ to mark $m$, $n$, that is, $m=qj'+l'$ and $n=qj+l$, where $q=2$ is the length of the primitive cell. With these notations,  we can reshape Eq.~\eqref{Diff} to
\begin{equation}
	\frac{z_m-z_n}{d} = (qj'+l')-(qj+l)-2\delta \sin{(2\pi\beta\frac{l+l'}{2}+\theta)}\sin{(2\pi\beta\frac{l'-l}{2})}.
\end{equation}
The eigenstates of the effective Hamiltonian in the real space satisfy
\begin{equation}
	\omega_{n,k}|\psi_{n,k}\rangle = H_{eff} |\psi_{n,k}\rangle,
	\label{6}
\end{equation}
with the Bloch states
\begin{equation}
	|\psi_{n,k}\rangle = \sum_{j,l}e^{ikqj}u_{n,k}(l)|qj+l\rangle,
\end{equation}
and the eigenvalues $\omega_{n,k}$.
Here, $k\in [-\pi/q,+\pi/q)$ is quasimomentum, $n$ is the band index, $j$ is the cell index, $l$ denotes the lattice index in a cell, and $u_{n,k}$ are periodic functions of Bloch states.
We multiply Eq.~\eqref{6} by $\langle qj'+l'|$ and obtain $u_{n,k}(l')$ satisfying
\begin{equation}
	(\omega_{n,k}-\omega_0)u_{n,k}(l') = -i\Gamma_0 \sum_{j,l} {e^{i\varphi|(qj'+l')-(qj+l)-2\delta\sin{(2\pi\beta\frac{l'+l}{2}+\theta)}\sin{(2\pi\beta\frac{l'-l}{2})}|} u_{n,k}(l) e^{ikq(j-j')}}.
\end{equation}
After a series of simplifications, we can obtain the equation below,
\begin{equation}
	\begin{split}
		&	(\omega_{n,k}-\omega_0)u_{n,k}(l') \nonumber \\
		&= i\Gamma_0 \frac{\sin{(kq)}}{\cos{(kq)}-\cos{(\varphi q)}} \sum_{l} \sin{\left\{ \varphi[l'-l-2\delta\sin{(2\pi\beta\frac{l'+l}{2}+\theta)}\sin{(2\pi\beta\frac{l'-l}{2})}] \right\}} u_{n,k}(l) \\
		&+\Gamma_0 \frac{\sin{(\varphi q)}}{\cos{(kq)}-\cos{(\varphi q)}} \sum_{l} \cos{\left\{ \varphi[l'-l-2\delta\sin{(2\pi\beta\frac{l'+l}{2}+\theta)}\sin{(2\pi\beta\frac{l'-l}{2})}] \right\} } u_{n,k}(l) \\
		&+\Gamma_0 \sum_{l'>l} \sin{\left\{ \varphi[l'-l-2\delta\sin{(2\pi\beta\frac{l'+l}{2}+\theta)}\sin{(2\pi\beta\frac{l'-l}{2})}] \right\} } u_{n,k}(l) \\
		&-\Gamma_0 \sum_{l'<l} \sin{\left\{ \varphi[l'-l-2\delta\sin{(2\pi\beta\frac{l'+l}{2}+\theta)}\sin{(2\pi\beta\frac{l'-l}{2})}] \right\} } u_{n,k}(l).
	\end{split}
\end{equation}
Finally, we have $(\omega_{n,k}-\omega_0)/\Gamma_0 u_{n,k}(l') = \sum_{l=1}H_k(l,l')u_{n,k}(l)$ with the elements of the Bloch Hamiltonian in the momentum space,
\begin{equation}
	\begin{split}
		H_k(l,l') = \sin{\big(\varphi\frac{|z_{l'}-z_l|}{d}\big)} + \frac{\sin{(\varphi q)} \cos{\big(\varphi\frac{|z_{l'}-z_l|}{d}\big)}}{\cos{(kq)}-\cos{(\varphi q)}}
		+\frac{i\sin{(k q)} \sin{\big(\varphi\frac{|z_{l'}-z_l|}{d}\big)}}{\cos{(kq)}-\cos{(\varphi q)}}
	\end{split}, \label{BlochMatrix}
\end{equation}
where the Hermitian matrix $H_k=H_k^\dagger$.
This means that the energy band of the excitation becomes real because there is no way for the excitation to escape from an infinite array.

In the non-Markov case, we need to solve nonlinear eigenvalue problem,
\begin{equation}
	H_k(\omega_{n,k})| u_{n,k}\rangle=\omega_{n,k}|u_{n,k}\rangle, \label{NonLinearNonMarkov}
\end{equation}
where the matrix elements of Hamiltonian are functions of eigenvalue $\omega=\omega_{n,k}$.
Anyway, there is no obstacle to treating the problem of the $2\times 2$ matrix in principle.
We can sweep the frequency to find the solution satisfying Eq.~\eqref{NonLinearNonMarkov} in a brute-force way.

To further simplify this problem, we consider the Markov approximation by substituting $\varphi = \omega_0 d/c$ into Hamiltonian elements.
From Eq.~\eqref{BlochMatrix}, we can deduce that the energy band is diverged at $k q=\pm \varphi q+2n\pi$, which originates from the Markov approximation of an infinite array.
However, the inverse energies $\bar \omega_{n,k}=(\omega_{n,k}-\omega_0)^{-1}$ are always  continuous functions of quasimomentum.

\subsection{Non-Markov energy band obtained via transfer matrix method}
In this subsection, we alternatively show how to derive energy band and Bloch functions by using the transfer matrix method without making the Markov approximation. 

We first show how to derive the transfer matrix for the transmission and reflection coefficients around the $j$th atom~\cite{dinc2019exactS}.
We need to solve the eigenvalue problem, $H_R |{E_{\omega}}\rangle  = {E_{{\omega}}}|{E_{{\omega}}}\rangle $ with $E_{\omega}=\hbar \omega$ and 
\begin{equation}
	|{E_{\omega}}\rangle  = \sum\limits_j^{} {{e_j}b_j^\dag |0\rangle }  + \int {dz{\phi _L}(z)a_L^\dag (z)|0\rangle }  + \int {dz{\phi _R}(z)a_R^\dag (z)|0\rangle }.  \label{Ek0}
\end{equation}
Here, $e_j$ is the probability amplitude of the excitation at the $j$th atom. $\varphi_R(z)$ and $\varphi_L(z)$ are the probability amplitudes for the rightward- and leftward-propagating photon at the position $z$, which are respectively given by
\begin{eqnarray} \label{Coefficients}
	{\phi _R}(z) &=& \frac{{{e^{i{\omega/c} z}}}}{{\sqrt {2\pi } }}\left[ {... + {t_1^2}\theta (z - {z_1})\theta ({z_2} - z) + {t_2^3}\theta (z - {z_2})\theta ({z_3} - z) + ...} \right], \\ 
	{\phi _L}(z) &=& \frac{{{e^{ - i{\omega/c} z}}}}{{\sqrt {2\pi } }}\left[ {... + {r_{1}^2}\theta (z - {z_1})\theta ({z_2} - z)  + {r_{2}^3}\theta (z - {z_2})\theta ({z_3} - z)+... } \right], \nonumber
\end{eqnarray}
where $t_j^{j+1}$ ($r_j^{j+1}$) is the transmission (reflection) coefficient between the $j$th and $(j+1)$th atoms.
By substituting Eq.~(\ref{Ek0}) into the Schr\"odinger equation with real-space Hamiltonian~\eqref{RealSpaceHam}, we can obtain
\begin{eqnarray} \label{Eigens}
	&&	\big( { - i\hbar c\frac{d}{{dz}} - {E_{\omega}}} \big){\phi _R}(z) + \hbar g\sum\limits_j^{} {\delta (z - {z_j}){e_j}}  = 0; \nonumber \\
	&&	\big( {i\hbar c\frac{d}{{dz}} - {E_{\omega}}} \big){\phi _L}(z) + \hbar g\sum\limits_j^{} {\delta (z - {z_j}){e_j}}  = 0; \nonumber\\
	&&	(\hbar {\omega _0} - {E_{\omega}}){e_j} + \hbar g\left[ {{\phi _L}({z_j}) + {\phi _R}({z_j})} \right] = 0.
\end{eqnarray}
Focusing on the $j$th atom ($z=z_j$), we substitute Eq.~(\ref{Coefficients}) into Eq.~(\ref{Eigens}) and can further obtain
\begin{eqnarray} \label{Eigenequations}
	&&	- i\frac{{{e^{i{\omega/c}{z_j}}}}}{{\sqrt {2\pi } }}\left( { - {t_{j - 1}^j} + {t_{j}^{j + 1}}} \right) + \frac{g}{c}{e_j} = 0; \nonumber \\
	&&	i\frac{{{e^{ - i{\omega/c}{z_j}}}}}{{\sqrt {2\pi } }}\left( { - {r_{j - 1}^j} + {r_{j}^{j + 1}}} \right) + \frac{g}{c}{e_j} = 0; \nonumber \\
	&&	\frac{{\omega}-{\omega _0} }{g} {e_j}=  {\frac{{{e^{i{\omega/c}{z_j}}}}}{{\sqrt {2\pi } }}\frac{{{t_{j-1}^j} + t_{j}^{j+1}}}{2} + \frac{{{e^{ - i{\omega/c}{z_j}}}}}{{\sqrt {2\pi } }}\frac{{{r_{j - 1}^j} + r_{j}^{j+1}}}{2}}. 
\end{eqnarray} 
By eliminating $e_j$ in Eq.~\eqref{Eigenequations}, we can build a relation between the reflection and transmission coefficients, 
\begin{eqnarray}
	\left( {\begin{array}{*{20}{c}}
			{t_j^{j + 1}}\\
			{r_j^{j + 1}}
	\end{array}} \right)&&=
	\left( {\begin{array}{*{20}{c}}
			{ - \left( {1 - f_{{\omega}}} \right)}&{f_{{\omega}}{e^{ - i2{\omega/c}{z_j}}}}\\
			{f_{{\omega}}{e^{2i{\omega/c} {z_j}}}}&{\left( {f_{{\omega}} + 1} \right)}
	\end{array}} \right)^{-1} \left( {\begin{array}{*{20}{c}}
			{ - 1 - f_{{\omega}}}&{ - f_{{\omega}}{e^{ - i2{\omega/c}{z_j}}}}\\
			{ - f_{\omega}{e^{2i{\omega/c}{z_j}}}}&{\left( {1 - f_{\omega}} \right)}
	\end{array}} \right)\left( {\begin{array}{*{20}{c}}
			{t_{j - 1}^j}\\
			{r_{j - 1}^j}
	\end{array}} \right)\\ \nonumber 
	&&	=M_j \left( {\begin{array}{*{20}{c}}
			{t_{j - 1}^j}\\
			{r_{j - 1}^j}
	\end{array}} \right),
\end{eqnarray}
with
\begin{equation}
	M_j=\left( {\begin{array}{*{20}{c}}
			{ - {{\left( {f_{\omega} + 1} \right)}^2} + f{{({\omega})}^2}}&{ - 2f_{\omega}{e^{ - i2{\omega/c}{z_j}}}}\\
			{2f_{\omega}{e^{2i{\omega/c}{z_j}}}}&{f_{\omega}^2 - {{\left( {1 - f_{\omega}} \right)}^2}}
	\end{array}} \right).
\end{equation}
Here, $f_{\omega} = {{i{\Gamma _0}}}/{{(2{\omega _0} -2\omega)}}$. 
With the transfer matrix $M_j$, we can build a relation between the reflection and transmission coefficients  $(t_{j-1}^{j},r_{j-1}^j)$ and $(t_{j+1}^{j+2},r_{j+1}^{j+2})$,
\begin{equation}
	\left( {\begin{array}{*{20}{c}}
			{t_{j+1}^{j + 2}}\\
			{r_{j+1}^{j + 2}}
	\end{array}} \right)= 	M_{j+1} M_j 	\left( {\begin{array}{*{20}{c}}
			{t_{j-1}^{j }}\\
			{r_{j-1}^{j }}
	\end{array}} \right)= 	T(\omega)	\left( {\begin{array}{*{20}{c}}
			{t_{j-1}^{j }}\\
			{r_{j-1}^{j }}
	\end{array}} \right), \label{Transfer2}
\end{equation}
where $T(\omega)$ is a $2\times 2$ matrix whose elements are functions of $\omega$.
According to the Bloch theorem, the reflection and transmission amplitudes also satisfy the relation,
\begin{equation}
	\left( {\begin{array}{*{20}{c}}
			{t_{j+1}^{j + 2}e^{i\omega/c z_{j+2}}}\\
			{r_{j+1}^{j + 2}e^{-i\omega/c z_{j+2}}}
	\end{array}} \right)= 	e^{ik q}	\left( {\begin{array}{*{20}{c}}
			{t_{j-1}^{j }e^{i\omega/c z_j}}\\
			{r_{j-1}^{j }e^{-i\omega/c z_j}}
	\end{array}} \right). \label{Bloch2}
\end{equation}
Combining Eqs.~\eqref{Transfer2} and \eqref{Bloch2}, we can obtain the following equation,
\begin{equation}
	T(\omega)	\left( {\begin{array}{*{20}{c}}
			{t_{j-1}^{j }}\\
			{r_{j-1}^{j }}
	\end{array}} \right)=\left( {\begin{array}{*{20}{c}}
			{ e^{i(k-\omega/c)q}}&{ 0}\\
			{0}&{e^{i (k+\omega/c) q}}
	\end{array}} \right) 	\left( {\begin{array}{*{20}{c}}
			{t_{j-1}^{j }}\\
			{r_{j-1}^{j } }
	\end{array}} \right).
\end{equation}
Because the determinant of $T$ is $1$, we can obtain a relation
\begin{equation}
	\cos(kq)=\frac{1}{2}[T_{11}(\omega)e^{i\omega/cq}+T_{22}(\omega)e^{-i\omega/cq}]. \label{Coskq}
\end{equation}
We can analytically obtain the matrix elements $T_{11}(\omega)$ and $T_{22}(\omega)$, and simplify Eq.~\eqref{Coskq} as
\begin{equation}
	\cos(kq)=	\frac{\Gamma_0^2}{(\omega_0-\omega)^2}\cos[k(z_3+z_1-2z_2)]+\left[1-\frac{\Gamma_0^2}{(\omega_0-\omega)^2}\right]\cos(kq)-\frac{2\Gamma_0}{\omega_0-\omega}\sin(kq).
\end{equation} 
We can numerically find the eigenfrequencies by solving the above equation. 

\subsection{Comparison between the Markov and non-Markov cases}
\begin{figure}[!htp]
	\center
	\includegraphics[width=0.99\textwidth]{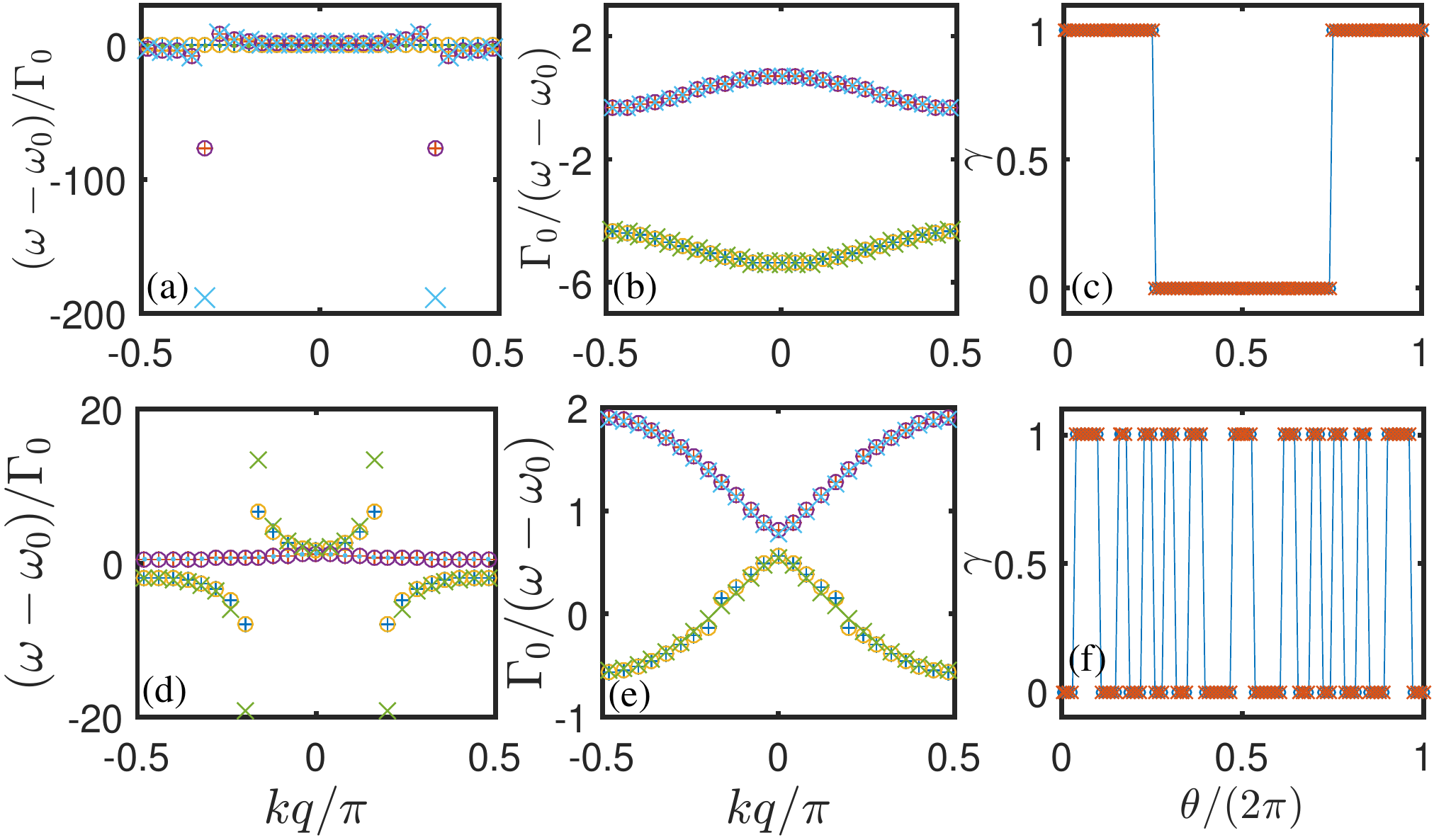}
	\caption{Comparison of (a,d) energy band, (b,e) inverse energy band, (c,f) Zak phase in the Markov and non-Markov cases. The parameters are chosen as $ c/d=100$ and $c/d=10$ in the top and bottom pannels. '$+$' denote the results obtained via transfer matrix method in the non-Markov case, '$\circ$' denote the results obtained by solving Eq.~\eqref{NonLinearNonMarkov} in the non-Markov case, and '$\times$' denote the results of the Markov case.  The parameters in calculations of energy bands and inverse energy bands are chosen as $g=1$, $\delta=0.4$, $N=25$, $\theta=0$, $\omega_0=100$, $c/d=100$ for the top pannel and $c/d=10$ for the bottom pannel. 
		Except for varying modulation phase, the parameters in calculations of Zak phase (c,f) correspond to those in (a,d), respectively. 
	}	\label{FigS0}
\end{figure}
In this subsection, we compare the energy band, inverse energy band, and Zak phase in the Markov and non-Markov cases; see Fig.~\ref{FigS0}.
The parameters are chosen as $c/d=100$ in Figs.~\ref{FigS0}(a,b) for weak non-Markov effect, and $c/d=10$ in Figs.~\ref{FigS0}(d,e) for  strong non-Markov effect, and the other common parameters are chosen as $g=1$, $\delta=0.4$, $N=25$, $\omega_0=100$, and $\theta=0$.
In the non-Markov case, we find that both the energy band and inverse energy band obtained via transfer matrix method (marked by '$+$') and solving Eq.~\eqref{NonLinearNonMarkov} (marked by '$\circ$') are exactly the same.
When the eigenfrequency is around $\omega_0$, the energy bands are almost the same in the Markov case (marked by '$\times$') and the non-Markov case (marked by '$+,\circ$').
This is because the Markov approximation [$d\ll c/(\omega-\omega_0)/N\sim \lambda/(2\pi N)$] works well around this energy region.
Around $kq=\pm \varphi q+2n\pi$, the eigenvalues tend to be divergent in the Markov case.
However, the eigenvalues in the non-Markov case are smaller than those in the Markov case.
We have also numerically verified that the eigenvalues in the non-Markov case do not tend to be divergent as the system size increases. 
The divergence in energy band comes from the Markov approximation of the infinite array.
Although energy bands in the non-Markov case overcome the divergence problem, the upper and lower branches are still disconnected in both the Markov and non-Markov cases.
The disconnection of the upper and lower branches hinders the application of conventional topological band theory in waveguide QED systems.
We find that the inverse energy band can fix this problem in both the Markov and non-Markov cases.
When the Markov approximation is valid in a small system with $N=50$, we find that the inverse energy bands in the Markov and non-Markov case are continuous functions of quasimomentum, and they are consistent with each other; see Fig.~\ref{FigS0}(b). 
When the non-Markov effect becomes stronger, the higher inverse energy bands in the Markov and non-Markov cases are consistent with each other, while the lower inverse energy band in the non-Markov case is slightly disconnected.
However, the lower inverse energy band can still be viewed as a band, because the inverse energies are close, and the Bloch eigenstates can be smoothly connected regardless of slight disconnection.
Furthermore, the Zak phases of inverse energy bands changing with modulation phase are exactly the same in both the Markov and non-Markov cases; see Fig.~\ref{FigS0}(c) for weak non-Markov effect and Fig.~\ref{FigS0}(f) for strong non-Markov effect. 
We should emphasize that the topological phase is not affected by the stronger non-Markov effect.

\section{Topological  phase  boundaries}
In this section, we derive the red solid lines denoting the topological phase boundaries in Fig.~2(a) of the main text.

The Bloch Hamiltonian can be expressed as
\begin{equation}
	H_{k} =
	\begin{pmatrix}
		a&b\\
		b^{*}&a
	\end{pmatrix},
\end{equation}
with $a = \frac{\sin{(\varphi q)}}{\cos{(kq)}-\cos{(\varphi q)}}$ and $b = \sin{(\varphi|\Delta z|)}+\frac{\sin{(\varphi q)} \cos{(\varphi|\Delta z|)}}{\cos{(kq)}-\cos{(\varphi q)}}+\frac{i\sin{(k q)} \sin{(\varphi|\Delta z|)}}{\cos{(kq)}-\cos{(\varphi q)}}$,
where the dimensionless distance is given by $\Delta z = (z_{2}-z_1)/d=1+2\delta\cos{\theta}$.
The inverse of the Bloch Hamiltonian can be written as
\begin{equation}
	\begin{split}
		H_{k}^{-1} 
		&=\frac{a}{a^2-|b|^2}
		\begin{pmatrix}
			1&0\\
			0&1
		\end{pmatrix}
		-\frac{1}{a^2-|b|^2}
		\begin{pmatrix}
			0&b\\
			b^{*}&0
		\end{pmatrix}.
	\end{split}
\end{equation}
The topological phase transition happens when the eigenvalues of $H_{k}^{-1}$ satisfy  $\bar\omega_{1,k}=\bar\omega_{2,k}$. The transition point turns out to satisfy $|b|^{2}=0$.
Moreover, the inverse energy band would be close at the quasimomentum $k=0$ or $k=\frac{\pi}{2}$.
In the case of $k=0$, $|b|^{2}=0$ indicates that
\begin{equation}
	\sin{(\varphi|\Delta z|)} =- \frac{\sin{(\varphi q)} \cos{(\varphi|\Delta z|)}}{1-\cos{(\varphi q)}}.
	\label{7}
\end{equation}
We substitute the expression $\Delta z = 1+2\delta\cos{\theta}$ and parameter $q=2$ into Eq.~\eqref{7} and obtain
\begin{equation}
	\varphi=\pm \frac{(2n-1)\pi}{4\delta\cos{\theta}} \ \text{or} \ \varphi=2 n\pi \quad (n =1,2...).
\end{equation}
Similarly, at $k=\frac{\pi}{2}$, $|b|^{2}=0$ indicates that
\begin{equation}
	\sin{(\varphi|\Delta z|)} = \frac{\sin{(\varphi q)} \cos{(\varphi|\Delta z|)}}{1+\cos{(\varphi q)}}.
\end{equation}
In a similar circumstance as $k=0$, at $k=\frac{\pi}{2}$ we can obtain the topological phase boundaries,
\begin{equation}
	\varphi=\pm \frac{n\pi}{2\delta\cos{\theta}} \ \text{or} \  \varphi=\pm\frac{(2n-1)\pi}{2}  \ \text{or} \ \theta=\frac{(2n-1)\pi}{2}
\end{equation}
with positive integers $n$.

\section{Flat  band and dark Wannier states}
In this section, we derive the parameters that support flat bands, and discuss the implication of the corresponding dark Wannier states.

When the inverse energy band is independent of $k$, $|b|^2$ should be a constant.
The constant magnitude of the complex number $b$ indicates that it should correspond to the value at $k=0$ as follows
\begin{eqnarray}
	&&\sin^2{(k q)} \sin^2{(\varphi|\Delta z|)} +	\left[\sin{(\varphi q)} \cos{(\varphi|\Delta z|)}+(\cos{(kq)}-\cos{(\varphi q)})\sin{(\varphi|\Delta z|)}\right]^2 \nonumber \\
	&&=\left\{\sin{(\varphi q)} \cos{(\varphi|\Delta z|)}+[1-\cos{(\varphi q)}]\sin{(\varphi|\Delta z|)}\right\}^2.
	\label{9}
\end{eqnarray}
Simplifying Eq.~\eqref{9}, we can obtain
\begin{equation}
	4\sin^2{\frac{kq}{2}} \sin{[(1+2\delta\cos{\theta}-q)\varphi]} \sin{[(1+2\delta\cos{\theta}) \varphi]}=0.
\end{equation}
Utilizing properties of trigonometric functions, we obtain the parametric equation about $\theta$ and $\varphi$,
\begin{equation}
	(-1+2\delta\cos{\theta})\varphi=\pm n\pi ~~\textrm{or}~~ (1+2\delta\cos{\theta}) \varphi=\pm n\pi
\end{equation}
with positive integers $n$.

\begin{figure}[!htp]
	\center
	\includegraphics[width=0.6\textwidth]{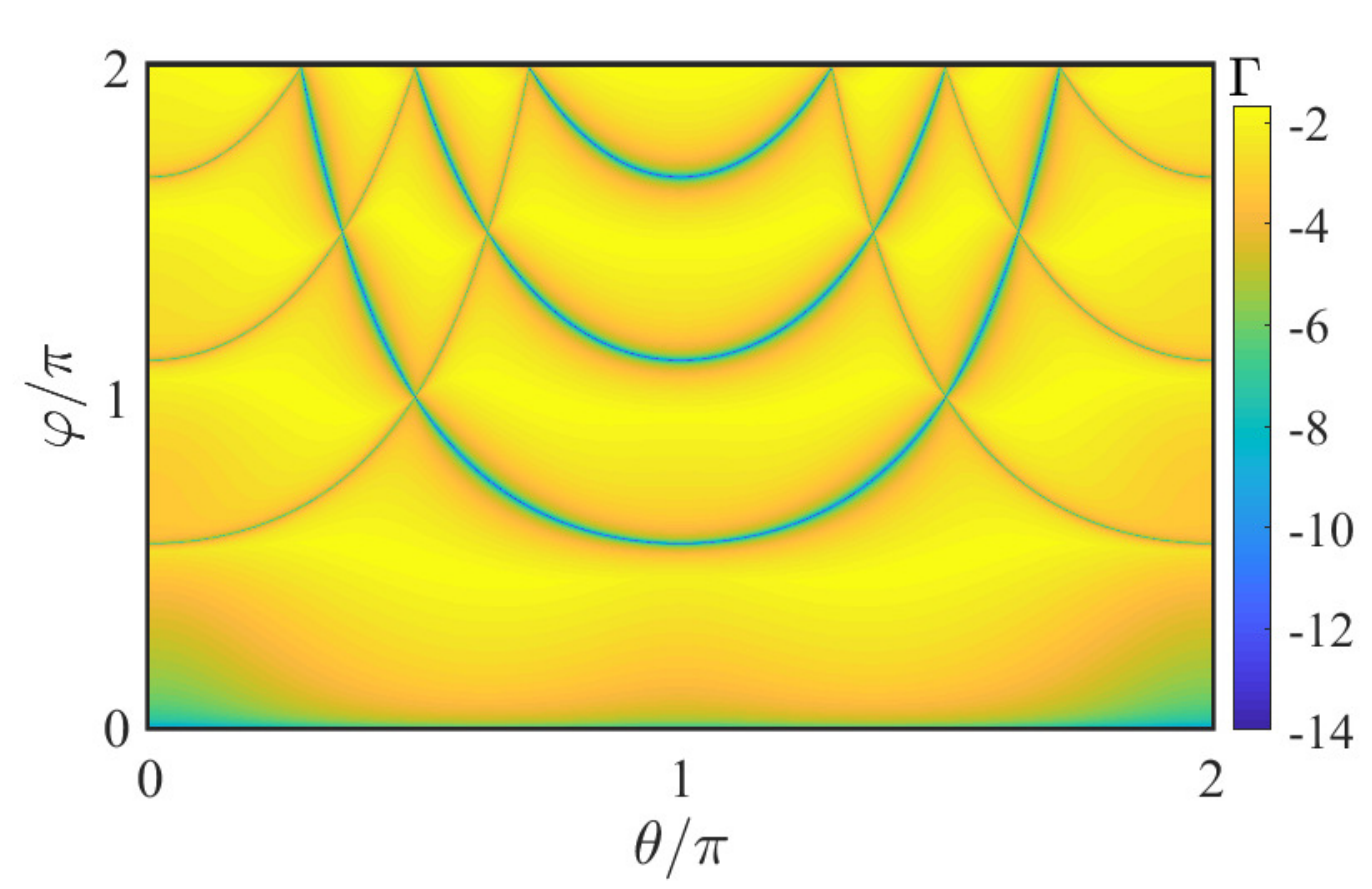}
	\caption{The decay rate of the most subradiant state as a function of modulation phase and phase constant. 
		The other parameters are chosen as $\Gamma_0=1$, $\delta=0.4$, and $N=4$.
	}	\label{FigS1}
\end{figure}
For the flat band, the group velocity is zero and the effective mass tends to infinite.
It means that an excitation will stay in the atomic array with zero decay rate.
Here, we show the decay rate [$\Gamma=-\text{Im}(\omega_n)$] of the most subradiant state as a function of $\theta$ and $\varphi$; see Fig.~\ref{FigS1}.
We can find the minimal decay rate in the parameters that support flat band, which are consistent with the analytical result.
The data reaches its limit of machine precision around $10^{-14}$, which can be viewed as zero decay rate.  

We need to emphasize the importance of the completely flat band, which supports a group of dark Wannier states.
These dark Wannier states form a decoherence-free subspace, in which dark Wannier states could be prepared, addressed, and accessed~\cite{zanner2022coherentS,Holzinger2022S}.
Such a structure could be a promising platform for quantum memory and quantum computation~\cite{Paulisch_2016S}.
%
%
The collective radiation of quantum emitters may significantly narrow the linewidth but maintain the resonant frequency~\cite{ChangDE2004S,Javanainen2017S,Javanainen2019S}.
The narrower linewidth is beneficial for improving the measurement precision of the resonant frequency.
More importantly, the undisturbed resonant frequency plays the same role as a magic wavelength in optical lattice clocks~\cite{campbell2017fermiS}.
Compared to individual atoms in optical lattice clocks, the collective radiation builds a coherent connection of all atoms and may be potentially used for the next generation of precision frequency measurement.

\section{Scale-free localized states}

\begin{figure}[!htp]
	\center
	\includegraphics[width=0.9\textwidth]{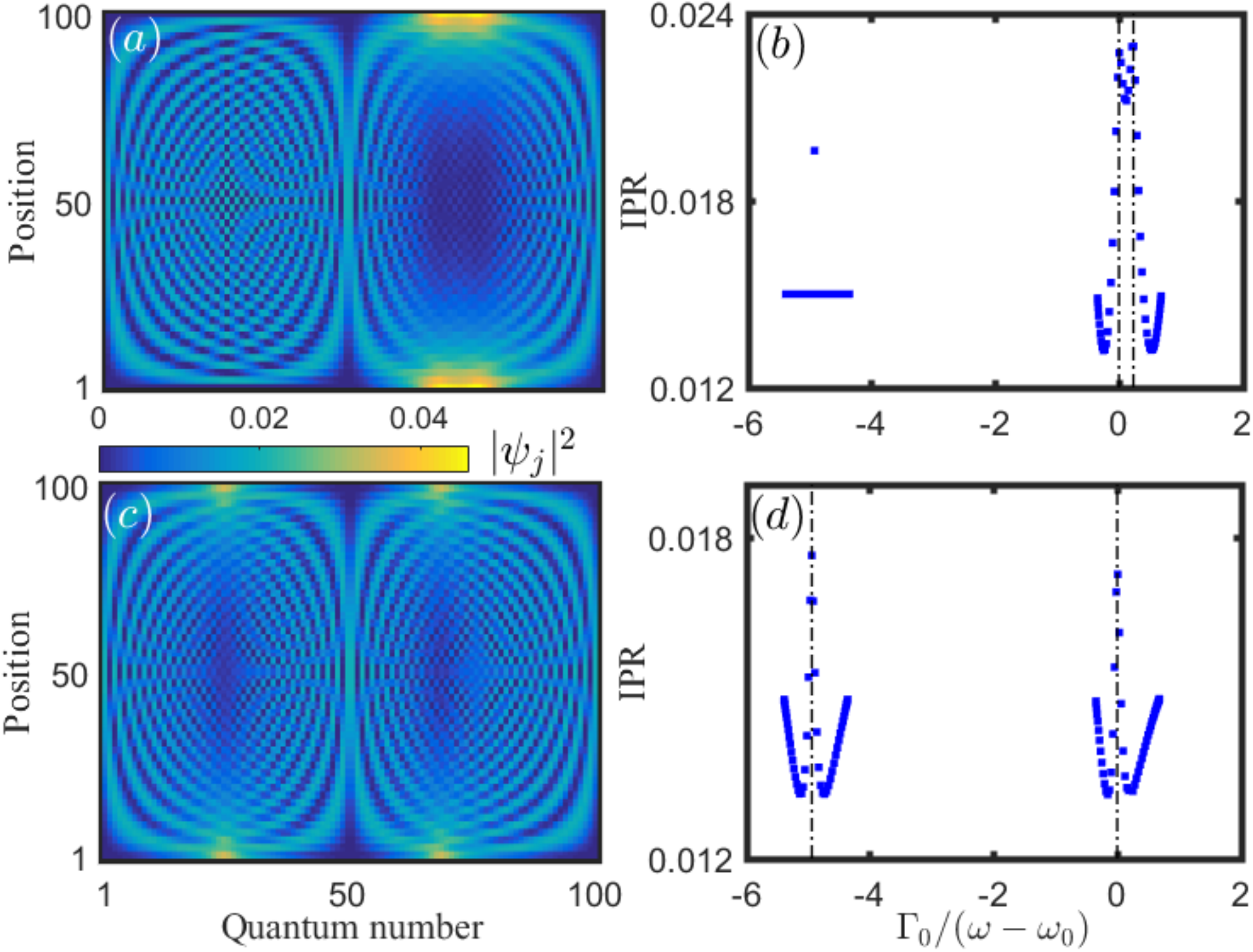}
	\caption{(a) and (c): density distribution of each eigenstate respectively with $\theta=0$ and $\theta=\pi$.
		The quantum number is ordered for increasing values of the inversed energy.
		(b) and (d) respectively correspond to the IPR of eigenstate in (a) and (c).
		The other parameters are chosen as $\phi=1$, $\Gamma_0=1$ and $\delta=0.4$.
		\label{fig:densitydistribution}}
\end{figure}

In this section, we explain why the radiative defects partially break the bulk-edge correspondence, analytically prove the topology-dependent distribution of scale-free localized states, and obtain the centered inverse energy of scale-free localized states.

We numerically find that $(H_{eff}-\omega_0)^{-1}$ corresponds to an SSH model with radiative defects at two endpoints in the following form:
\begin{equation}\label{SSHmodel}
	H_{SSH}=\sum_{j}\left[v c_{2j-1}^{\dag} c_{2j}+w c_{2j}^{\dag} c_{2j+1}+\text { h.c. }+C (n_{2j-1}+n_{2j}) \right]+D\left(n_{1}+n_{2N}\right).
\end{equation}
Here, $v$ and $w$ respectively denote the intra-cell and inter-cell hopping strengths, $D$ is a complex number, denoting the radiative defect at the boundaries, and $C$ is an energy constant.
According to the inverse relation, $H_{SSH}(H_{eff}-\omega_0)=I$ with $I$ being the identity matrix,  we find that  $v$, $w$, $C$ and $D$ satisfy
\begin{eqnarray}
	&&	-i\Gamma_0 D -i\Gamma_0 v e^{i\varphi|z_1-z_2|}=1, \nonumber \\
	&&  -i\Gamma_0 D e^{i\varphi|z_1-z_2|} -i\Gamma_0 v=0, \nonumber \\
	&&  -i\Gamma_0 v e^{i\varphi|z_1-z_2|} -i\Gamma_0 C -i\Gamma_0 w e^{i\varphi |z_2-z_3|}=1, \nonumber \\
	&&  -i \Gamma_0 v e^{i\varphi|z_1-z_3|}-i\Gamma_0 C e^{\varphi|z_2-z_3|}-i\Gamma_0 w=0.
\end{eqnarray}
Solving the above equations, we can analytically obtain the relation between the parameters of the SSH-type Hamiltonian and those of the original effective Hamiltonian,
\begin{eqnarray}
	v&=&\frac{1}{2\Gamma_0\sin(\varphi|z_2-z_1|/d)},\nonumber \\
	w&=&\frac{1}{2\Gamma_0\sin(\varphi|z_3-z_2|/d|)},\nonumber \\
	C&=&-\frac{1}{2\Gamma_0}\left[\cot(\varphi|z_2-z_1|/d)+\cot(\varphi|z_3-z_2|/d)\right],\nonumber \\
	D&=&\frac{i}{2\Gamma_0}-\frac{1}{2\Gamma_0}\cot(\varphi|z_2-z_1|/d). \label{Parameters}
\end{eqnarray} 
In the case of equal-spacing array, these parameters can return to those in Ref.~\cite{Poddubny2020S}. From the expression of $D$, we can clearly see that the SSH-type Hamiltonian is non-Hermitian with radiative decay at the boundaries. This is the main reason for the partial breakdown of the bulk-edge correspondence.

We show the density distribution and the inverse participation ratio (IPR) of each eigenstate for the topological phase ($\varphi=1,\theta=0$) and the trivial phase 
($\varphi=1,\theta=\pi$) in Fig.~\ref{fig:densitydistribution}.
Unlike topological edge states in the gap, Fig.~\ref{fig:densitydistribution} indicates that scale-free localized states exist in the upper inverse energy band for the topological phase while in both upper and lower inverse energy bands for the trivial one.
This means that topology plays a crucial role in the distribution of scale-free localized states. 
In the following, we will analytically demonstrate the key role of the topological phase.

To derive the eigenvalues of scale-free localized states, we consider a semi-infinite SSH-type model with an overall shift of the energy constant,
\begin{equation}\label{SSHmodelSemi}
	\tilde H_{SSH}=\sum_{j=1}^{N}\left(v c_{2j-1}^{\dag} c_{2j}+w c_{2j}^{\dag} c_{2j+1}+\text { h.c. } \right)+(D-C) n_{1},
\end{equation}
where $N$ tends to be infinite. We assume that the wavefunction of the scale-free localized state satisfies 
\begin{eqnarray}
	\psi_1&=&A, \nonumber \\
	\psi_{2j-1}&=&a_1 R e^{ikj}+a_2 S e^{-ikj}, \nonumber \\
	\psi_{2j}&=& a_1 S e^{ikj}+a_2 R e^{-ikj}, 
\end{eqnarray}
where $A$, $R$, and $S$ are parameters to be determined, and $a_1$, $a_2$ are rather arbitrary due to the degeneracy of eigenenergies at momenta $k$ and $-k$.
According to the Schr\"ordinger equation, $\tilde{H}_{SSH}|\psi\rangle=E|\psi\rangle$, we can obtain
the following equations around the boundary,
\begin{eqnarray} \label{Eqboundary}
	[E-(D-C)]A&=&v(a_1Se^{ik}+a_2 Re^{-ik}), \nonumber \\
	E(a_1 S e^{ik}+a_2 R e^{-ik})&=&vA+w(a_1 Re^{i2k}+a_2 S e^{-i2k}), 
\end{eqnarray}
and in the bulk,
\begin{eqnarray}
	E(a_1 R e^{ik(N-1)}+a_2 S e^{-ik(N-1)})	&=&w(a_1 Se^{ik(N-2)}+a_2 R e^{-ik(N-2)})+v(a_1 Se^{ik(N-1)}+a_2Re^{-ik(N-1)}).\nonumber \\
	E(a_1 S e^{ik(N-1)}+a_2 R e^{-ik(N-1)})	&=&v(a_1 Re^{ik(N-1)}+a_2 S e^{-ik(N-1)})+w(a_1 Re^{ikN}+a_2Se^{-ikN}). \nonumber \\
	\label{Bulk}
\end{eqnarray}

Combining Eqs.~\eqref{Eqboundary} and ~\eqref{Bulk}, we can obtain a relation
\begin{eqnarray}
	\frac{e^{-ik(N-2)}}{e^{ik(N-2)}}=\frac{E-\frac{v^2}{E-(D-C)}-w\frac{S}{R}e^{ik}}{E\frac{S}{R}-\frac{S}{R}\frac{v^2}{E-(D-C)}-we^{ik}}\frac{E\frac{S}{R}-v-we^{ik}}{E-v\frac{S}{R}-w\frac{S}{R}e^{-ik}}.
\end{eqnarray}
The decay behavior of scale-free localized states indicates that the quasimomentum $k$ is a complex number, so we denote $k=-i\rho+k_R$ with $\rho>0$ and $k_R$ being real.
When $N\rightarrow \infty$, the above equation tends to be zero, and consequently we can obtain
\begin{equation}
	E-\frac{v^2}{E-(D-C)}-w\frac{S}{R}e^{-\rho-ik_{R}}=0. \label{BulkRelation}
\end{equation}
Because the scale-free localized states have spatial decay rate inversely proportional to the system size~\cite{li2021impurityS,li2023scalefreeS}, $\rho=\alpha/N$, which tends to be zero in the limits of $N\rightarrow \infty$. 
Here, the ratio $S/R$ and the energy can be determined by Eq.~\eqref{Bulk}, which are respectively given by
\begin{eqnarray}
	\frac{S}{R}=\frac{E}{v+we^{-ik_R}}=\frac{v+we^{ik_R}}{E}, \label{RatioSR}
\end{eqnarray}
and 
\begin{equation}
	E=\pm\sqrt{v^2+w^2+2vw\cos(k_R)}. \label{EnergyB}
\end{equation}
Unlike conventional topological edge states in the band gap, we can find that in the limits of $N\rightarrow \infty$ the eigenvalues of scale-free localized states are in the bulk bands.

To further obtain the position of scale-free localized states in the bulk bands, we substitute Eqs.~\eqref{RatioSR} and \eqref{EnergyB} into Eq.~\eqref{BulkRelation} and obtain an equation for $k_R$ and $E$,
\begin{equation}
	\frac{E(D-C)}{w}=ve^{ik_R}+w.
\end{equation} 
According to Eq.~\eqref{Parameters}, we find $(D-C)/w=e^{i\varphi|z_3-z_2|/d}$ and the above equation is simplified as 
\begin{equation}
	E=e^{-i\varphi|z_3-z_2|}(ve^{ik_R}+w). \label{ECos}
\end{equation}
From Eq.~\eqref{ECos}, we can deduce that the eigenvalues of scale-free localized states can only appear in one band in the topological phase $|v|<|w|$ and can appear in two bands in the trivial phase $|v|\ge|w|$.

Furthermore, separating the real and imaginary parts of Eq.~\eqref{ECos}, we can obtain two equations,
\begin{eqnarray}
	&&	E=v\cos(k_R-\varphi|z_3-z_2|)+w\cos(\varphi|z_3-z_2|),\nonumber \\
	&&	v\sin(k_R-\varphi|z_3-z_2|)=-w\sin(\varphi|z_3-z_2|),
\end{eqnarray}
with which we can derive the centered positions of scale-free localized state in the inverse energy band; see  black dotted-dashed lines in Fig.~\ref{fig:densitydistribution} and the white dashed line in Fig.~2(c) of the main text.
We can see that the centered inverse energies are consistent with the peak positions of the IPRs.

\section{Extraction of topological phase via quantum walks}
\begin{figure}[!htp]
	\includegraphics[width=0.7\textwidth]{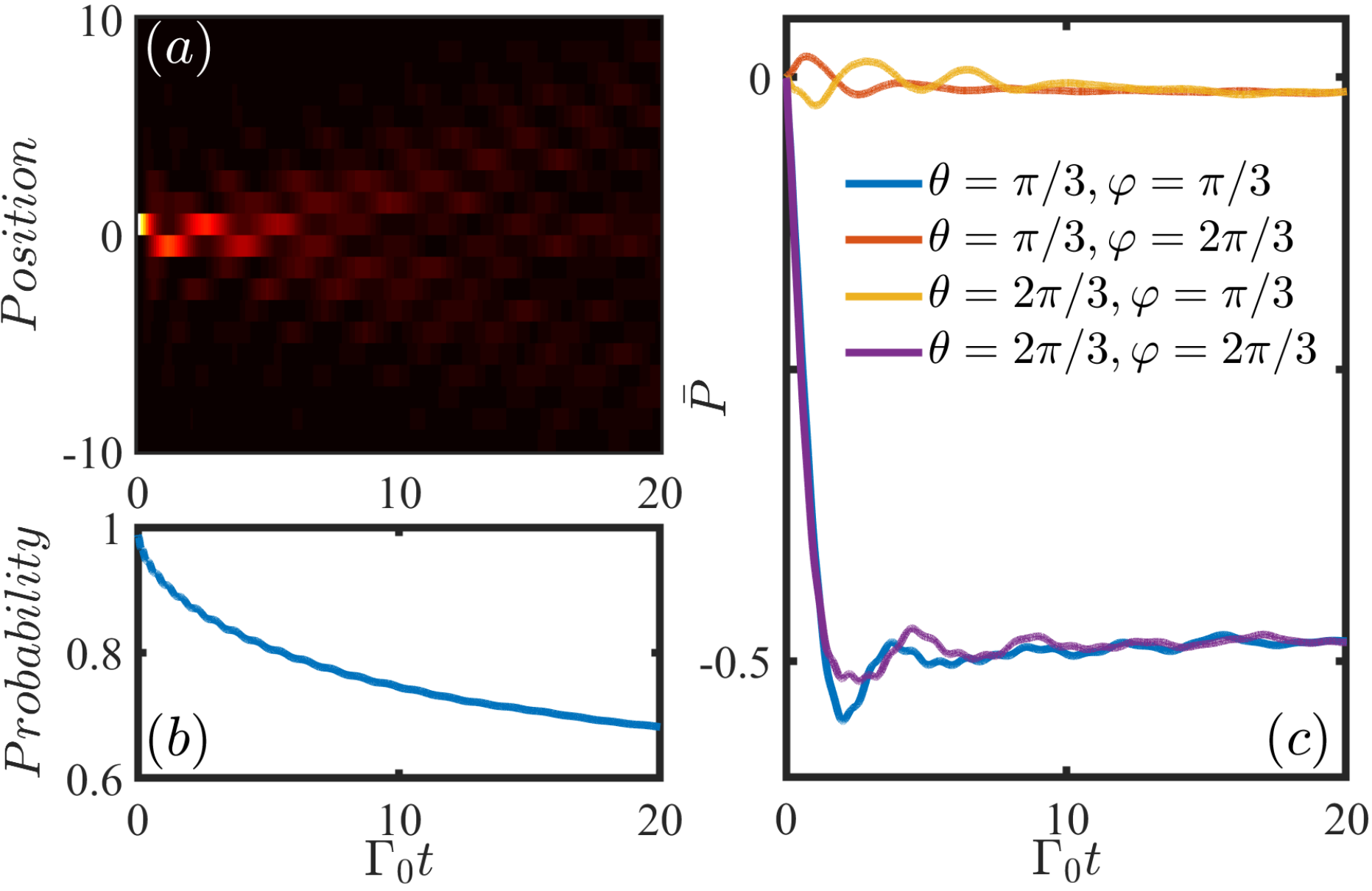}
	\centering
	\caption{(a) Quantum walks and (b) decay of excitation in topological nontrivial phase with ($\theta=\pi/3, \ \varphi=\pi/3$). (c) Mean cell positions in different topological phases. The other parameters are chosen as $\delta=0.4$ and $N=51$.
	}\label{FigS40}
\end{figure}
In the main text, we mention that we can further extract the Zak phase of the inverse energy band via long-time average of mean cell position in quantum walks.
Since edge state is no longer an evidence for topological phase, below we show how to extract the topological phase from the dynamics of an excitation.

The effective Hamiltonian for excitation has inversion symmetry and breaks chiral symmetry due to the long-range couplings.
Adopted the dynamical scheme in Refs.~\cite{Longhi18S,Jiao2021S}, we propose to prepare an excitation in the $N$th emmiter and observe the time average of the mean cell position of excitation undergoing quantum walks,
\begin{equation}
	\bar P(T)=\frac{1}{T}\int_0^T\sum\limits_{j=1}^N j(|\psi(2j-1,t)|^2+|\psi(2j,t)|^2)dt.
\end{equation}
Here, $\psi(l,t)$ is the probability amplitude of the excitation at the $l$th atom at time $t$, which is obtained by $|\psi(t)\rangle=\exp(-i H_{eff}t/\hbar)|\psi(0)\rangle$, and $T$ is the total time. The Zak phase is related to a winding number, $\mathcal W$, and hence the mean cell position,
\begin{equation}
	\gamma=\mathcal{W} \pi=-2 \pi \lim\limits_{T\rightarrow\infty} \bar P(T).
\end{equation}
However, in our system, we have to compromise that the time should be less than the lifetime of excitation and large enough for the topological phase to take effect. 
Given $\delta=0.4$, $\theta=\pi/3$, $\varphi=\pi/3$ and $N=51$, as time evolves, we show the probability distribution ($|\psi(j,t)|^2$) of an excitation in Fig.~\ref{FigS40}(a) and the total probability $\sum_j|\psi(j,t)|^2$ in Fig.~\ref{FigS40}(b).
Even with excitation decay, we can still approximately extract the $\pi$ Zak phase through $\bar P \approx -0.5$ and the $0$ Zak phase through $\bar P \approx 0$; see Fig.~\ref{FigS40}(c). 
For comparison, we also show the mean cell positions for the topological nontrivial phase with $\theta=2\pi/3$, $\varphi=2\pi/3$ (dark blue line) and trivial phases with $\theta=\pi/3$, $\varphi=2\pi/3$ (red line) and $\theta=2\pi/3$, $\varphi=\pi/3$ (yellow line).

\section{Green's function method}
In this section, we derive the Green's function method to give the reflection and transmission coeffecients in Eqs.~(4) and (5) of the main text.

First, we separate the Hamiltonian~\eqref{eq:H0} in momentum space into two parts, $H_M=H_0+V$, with the unperturbed term,
\begin{equation}
	H_0=\int_{-\infty}^{+\infty} \hbar\omega_{k}a(k)^{\dag}a(k)^{\vphantom{\dag}}dk
	+\sum\limits_{j}\hbar\omega_{0}b_{j}^{\dag}b_{j}^{\vphantom{\dag}},
\end{equation}
and the perturbed term
\begin{equation}
	V=\frac{\hbar g}{\sqrt{2\pi}}\sum\limits_{j}\int_{-\infty}^{+\infty}
	(b_{j}^{\dag}a{(k)}e^{i k z_{j}}+b_{j}a{(k)}^{\dag}e^{-i k z_{j}})dk.
\end{equation}
The scattering problem can be solved by the famous Lippmann-Schwinger equation,
\begin{equation}
	|{\psi _{out}}\rangle  = |{\psi _{in}}\rangle  + \frac{1}{{E - {H_0}}}V|{\psi _{out}}\rangle, \label{LSEqn}
\end{equation}
where $|\psi_{in}\rangle=a(\kappa)^\dagger|0\rangle$ indicates that an input photon with a wave vector $\kappa>0$ and frequency $\omega=c|\kappa|$ is propagating rightward.
The output state is given by
\begin{equation}
	|{\psi _{out}}\rangle  = \sum\limits_{{j}}^{} {{Q_{{j}}}b_{{j}}^\dag |0\rangle }  + \int {{P_{{k}}}a(k) ^\dag|0\rangle dk },
\end{equation}
where $P_{k}$ and $Q_j$ are the probability amplitudes of the photon with momentum $k$ and excitation at the $j$th atom, respectively.
Substituting the input and output states into Eq.~\eqref{LSEqn}, we can obtain the relation between $P_k$ and $Q_j$,
\begin{eqnarray}
	{P_k} &=& \delta ({k,{\kappa}}) + \frac{g}{{\sqrt {2\pi } }}\sum\limits_j^{} {\frac{{{e^{ - ik{z_j}}}}}{{\omega  - c|k|}}{Q_j}} ; \label{Photon} \\
	{Q_j} &=& \frac{g}{{\sqrt {2\pi } }}\int {\frac{{{e^{ik{z_j}}}}}{{\omega  - {\omega _0}}}{P_k}dk}.
\end{eqnarray}
Combining the above equations, we can obtain the relations for the motion of an excitation,
\begin{equation}
	{Q_j} = \frac{g}{{\sqrt {2\pi } }}\frac{{{e^{i\kappa {z_j}}}}}{{\omega  - {\omega _0}}} - i\Gamma_0\sum\limits_{j'}^{} {\frac{{{e^{i\omega /c|{z_j} - {z_{j'}}|}}}}{{\omega  - {\omega _0}}}{Q_{j'}}}, \label{excitation}
\end{equation}
where we have used a formula
\begin{equation}
	\frac{g^2}{2\pi}\int \frac{e^{ik(z_j-z_{j'})}}{\omega-c|k|}dk=-i \frac{g^2}{c}e^{i\omega/c|z_j-z_{j'}|}.
\end{equation}
The propagation of an excitation in the atomic array is governed by the Green's function, which is determined by
\begin{equation}
	{G^{ - 1}}(\omega ) = \omega -H_{eff}. \label{greendefinition}
\end{equation}
Combining Eqs.~\eqref{excitation} and \eqref{greendefinition}, the excitation amplitudes $Q_j$ are given by
\begin{equation}
	Q_j=\frac{g}{\sqrt{2\pi}}\sum_{j'} {G_{j,j'} (\omega) e^{i\kappa z_{j'}}}. \label{Qj}
\end{equation}
Substituting Eq.~\eqref{Qj} into Eq.~\eqref{Photon}, we can obtain the output field distribution
\begin{equation}
	P_k=\delta(k,\kappa)+\frac{g^2}{2\pi}\sum_{j,j'}G_{j,j'}(\omega_{\kappa})\frac{e^{-ik z_j+i\kappa z_{j'}}}{\omega-c|k|}.
\end{equation}
The reflection coefficient is given by
\begin{equation}
	r_{\kappa}=\int_{-\infty}^{0} P_k dk =-i\Gamma_0 \sum_{j,j'}G_{j,j'}(\omega_{\kappa})e^{i\omega_{\kappa}/c (z_j+z_{j'})}, \label{reflectionk}
\end{equation}
and
the transmission coefficient is given by
\begin{equation}
	t_{\kappa} =\int_{0}^{+\infty} P_k dk = 1 - i{\Gamma _0}\sum\limits_{j,j'}^{} {{{G_{j,j'}}(\omega_k ) e^{i\omega_k /c({z_{j'}} - {z_j})}}}, \label{transmission}
\end{equation}
and they also satisfy $|r_{\kappa}|^2+|t_{\kappa}|^2=1$. %

Because the reflection can be immediately obtained once the transmission is known, we focus on the reflection of a single photon.
To make it clear, we expand the Green function in terms of the eigenvalues $\{\omega_n\}$ and the eigenstates $|\{\psi_n\}\rangle$ of the effective Hamiltonian,
\begin{equation}
	G_{j,j'}(\omega_{\kappa})=\sum\limits_{n}\frac{\psi_n(j)\psi_n(j')}{\omega_{\kappa}-\omega_n}, \label{Green}
\end{equation}
where the eigenstate $|\psi_n\rangle$ has been normalized through $\psi_n(j)=\psi_n(j)/\sqrt{\sum_j \psi_n(j)^2}$.
By combining Eqs.~\eqref{Green} and \eqref{reflectionk}, we can obtain the following result
\begin{equation}
	r_{\kappa} = -i{\Gamma _0}\sum\limits_{j,j',n}^{} {{e^{i\omega_{\kappa} /c({z_{j'}} + {z_j})}}\frac{\psi_n(j)\psi_n(j')}{\omega_{\kappa}-\omega_n}}, \label{transmission1}
\end{equation}
When the frequency of the photon is in resonance with the energy of an excitation state [i.e., $\omega_{\kappa}=\textrm{Re}(\omega_n)$],
then the denominator only takes the value of the imaginary part of the eigenvalue $\textrm{Im}(\omega_n)$.
If the eigenstate $|\psi_n\rangle$ is a subradiant state with radiative decay rate $-\textrm{Im}(\omega_n)\ll \Gamma_0$, then the reflection is mainly determined by the properties of the $n$th eigenstate.
Hence, we can neglect the summation over the quantum number in Eq.~\eqref{transmission1} and obtain the reflection coefficient
\begin{equation}
	r_{\kappa} \approx  {\Gamma _0}\sum\limits_{j,j'}^{} {{e^{i\omega_{\kappa} /c({z_{j'}} + {z_j})}}\frac{\psi_n(j)\psi_n(j')}{\textrm{Im}(\omega_n)}}. \label{transmission}
\end{equation}
Thus, reflection may give information about topological properties of subradiant states.

\section{Winding number of scattering textures}
\begin{figure}[!htp]
	\center
	\includegraphics[width=1\textwidth]{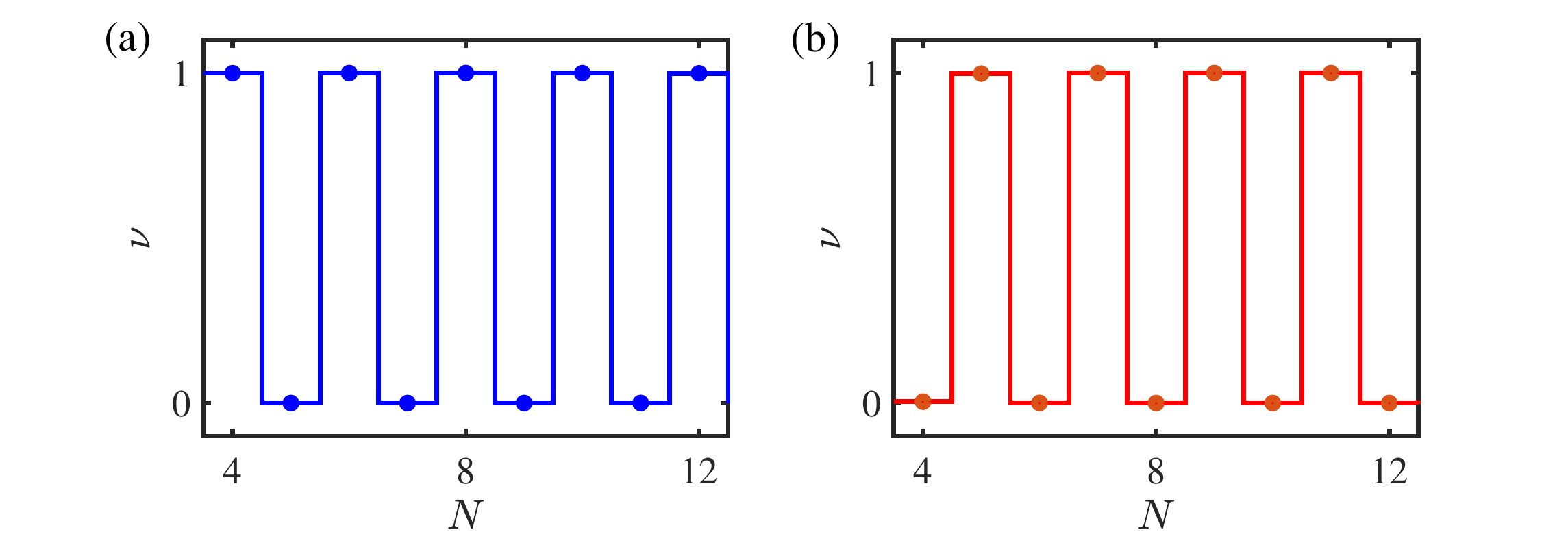}
	\caption{Winding number of scattering texture as a function of cell number in (a) topological nontrivial phase and (b) trivial phase.
		The parameters are chosen as $g=1$, $\Gamma_0=0.01$, $\omega_0=c/d=100$, $\delta=0.4$, $\varphi=1$, $\theta=0$ and $\theta=\pi$ for topological and trivial phases, respectively.
		\label{FigSup4}}
\end{figure}
In this section, we show the summarized results in the Table~I of the main text.

Without loss of generality, we choose the two typical sets of parameters in the topological nontrivial phase $(\varphi\approx 1, \ \theta=0)$ and trival phase $(\varphi\approx 1, \ \theta=\pi)$, and calculate the winding number of scattering textures in the lower inverse band of subradiant states; see Fig.~\ref{FigSup4}.
The other parameters are chosen as $g=1$, $\Gamma_0=0.01$, and $\omega_0=c/d=100$.
When the number of cells is not less than $4$, the winding number of scattering textures is precisely quantized, either $0$ or $1$.
No matter for topological nontrivial phase or trivial phase, the winding number strongly depends on the even-odd number of cells. 
For the topological nontrivial phase, the winding number is $1$ for the even cell number and $0$ for the odd cell number.
For the trivial phase, the winding number is $0$ for the even cell number and $1$ for the odd cell number.
This is the first time that we find that the winding number of scattering textures is affected by both the topological phase of excitation and the even-odd effect of cell number.

\section{Phase shift due to subradiant states}

\begin{figure}[!htp]
	\center
	\includegraphics[width=1\textwidth]{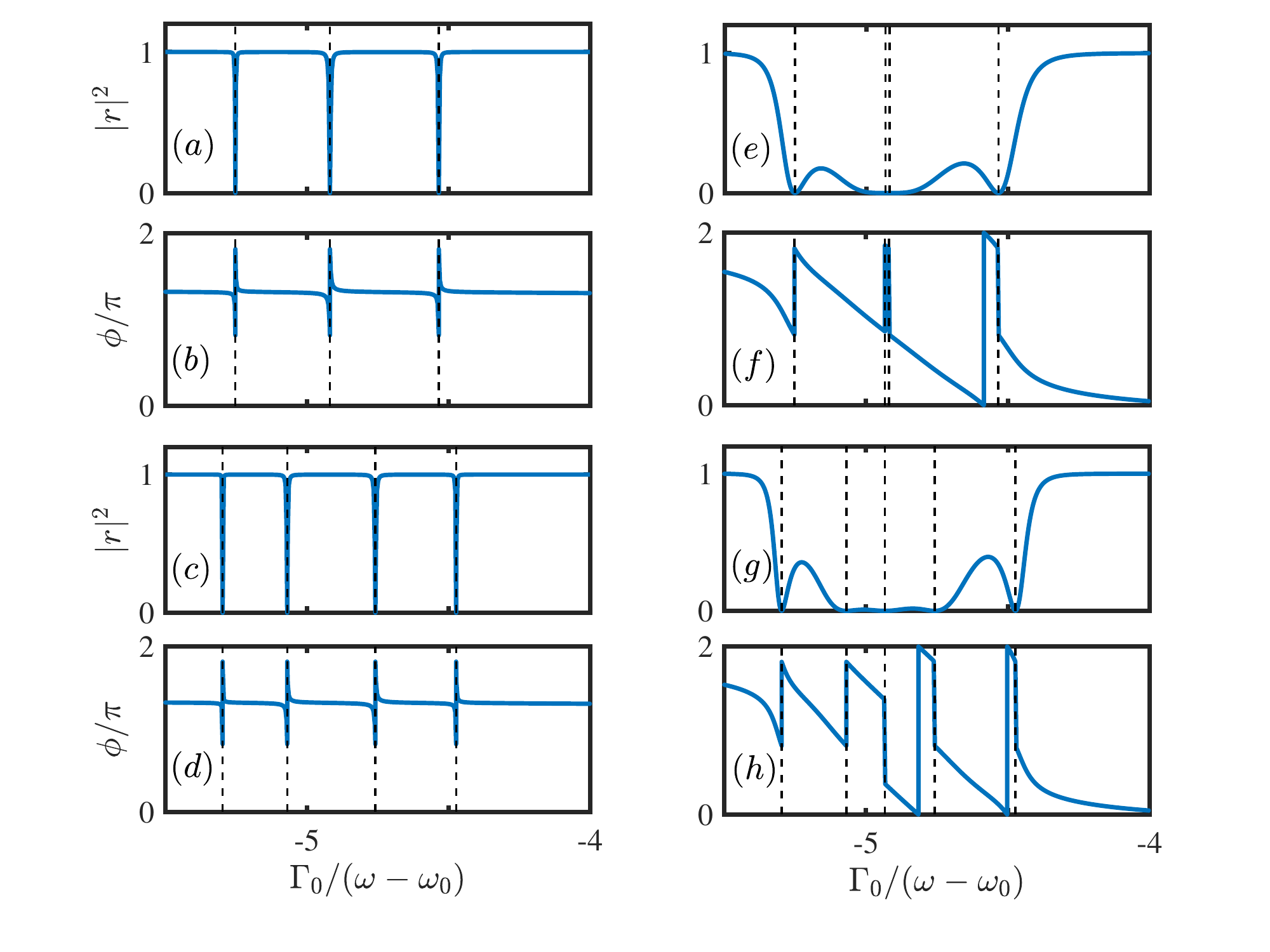}
	\caption{ Reflection and phase of reflection as functions of inverse energy. The left and right panels correspond to topological nontrivial phase $(\varphi\approx 1, \ \theta=0)$ and trival phase $(\varphi\approx 1, \ \theta=\pi)$, respectively. $(a,b, e, f)$, $(c,d,g,h)$ show the Reflection and phase of reflection in $4$ and $5$ cells, respectively. The dash lines indicate the inverse energies for the dip and $\pi$ phase shift, which are around the energies of subradiant states.
		The other parameters are chosen as $g=1$, $\delta=0.4$, $\Gamma_0=0.01$ and $\omega_0=c/d=100$.
		\label{FigS5}}
\end{figure}
In this section, we numerically verify that there exists a $\pi$ phase shift due to the subradiant states.

We calculate reflection and phase of reflection by sweeping the inverse energy through the lower inverse energy band of the subradiant states; see Fig.~\ref{FigS5}.
For the case of the nontrivial topological phase (left panel of Fig.~\ref{FigS5}), there are $(N-1)$ subradiant states for $N$ cells.
When the frequency of a photon is in resonance with the subradiant states, the photon becomes completely transmission, and there exist dips at the inverse energy of the subradiant states.
In Figs.~\ref{FigS5}(a) and (c), the number of dips in the lower inverse energy band is $3$ and $4$, corresponding to $4$ and $5$ cells.
We can also find that there exists a phase shift $\pi$ at each resonant point in Figs.~\ref{FigS5}(b) and (d), which are calculated with the same parameters as Figs.~\ref{FigS5}(a) and (c), respectively.
For the case of the trivial phase (right panel of Fig.~\ref{FigS5}), there are $N$ subradiant states for $N$ cells.
There are $4$ and $5$ dips in the reflection spectrum with $4$ and $5$ cells; see Figs.~\ref{FigS5}(e) and (g).
Because the decay rates of the subradiant states are relatively larger and comparable to the energy difference between the subradiant states, the reflection cannot simply be attributed to individual subradiant states but rather to the collective effect of the nearby subradiant states.
Hence, the positions of the dips slightly depart from the inverse energies of individual subradiant states.
However, the $\pi$ phase shifts are also accompanied by the dips; see Figs.~\ref{FigS5}(f) and (h).
By sweeping through the lower inverse energy band of the subradiant states, we can find that there are $(N-1)$ and $N$ times of $\pi$ phase shifts for topological nontrivial and trivial phases, respectively.
Since the sum of $\pi$ phase shifts depends on both the topological phase of excitation and the even-odd number of cells, similar to the winding number of scattering textures, the $\pi$ phase shift induced by subradiant states and the winding number should have an intimate relation.
We can attribute the winding number of scattering textures to the sum of $\pi$ phase shifts.

%

\end{document}